\documentclass[aps,showpacs,preprintnumbers,amsmath,amssymb]{revtex4}

\UseRawInputEncoding
\usepackage{balance}
\usepackage{hyperref} 
\hypersetup{colorlinks=true, linkcolor=blue, citecolor=blue} 
\usepackage{dcolumn}
\usepackage[dvips]{epsfig}

\usepackage{float}
\usepackage{graphicx}
\usepackage{epstopdf}
\usepackage{graphicx}
\usepackage{epstopdf}
\usepackage{color}
\usepackage{amsmath,amssymb,amsfonts}

\begin{document}
\baselineskip=0.8 cm

\title{Gravitational waves from equatorially eccentric  extreme mass ratio inspirals around swirling-Kerr black holes }
\author{Yuhang Gu$^{1}$,
    Songbai Chen$^{1,2}$\footnote{Corresponding author: csb3752@hunnu.edu.cn},
    Jiliang Jing$^{1,2}$ \footnote{jljing@hunnu.edu.cn}}

\affiliation{$^1$Department of Physics, Institute of Interdisciplinary Studies, Hunan Research Center of the Basic Discipline for Quantum Effects and Quantum Technologies, Key Laboratory of Low Dimensional Quantum Structures
    and Quantum Control of Ministry of Education, Synergetic Innovation Center for Quantum Effects and Applications, Hunan
    Normal University,  Changsha, Hunan 410081, People's Republic of China
    \\
    $ ^2$Center for Gravitation and Cosmology, College of Physical Science and Technology, Yangzhou University, Yangzhou 225009, People's Republic of China}

\begin{abstract}
\baselineskip=0.6 cm
\begin{center}
{\bf Abstract}
\end{center}
The swirling-Kerr black hole is a novel solution of vacuum general relativity and has an extra swirling parameter
characterizing the rotation of spacetime background. We have studied the  gravitational waves generated by extreme mass ratio inspirals (EMRIs) along eccentric orbits on equatorial plane in this novel swirling spacetime.  Our findings indicate that this swirling parameter leads to a delayed phase shift in the gravitational waveforms. Furthermore, we have investigated effects of the swirling parameter on the potential issue of waveform confusion caused by the orbital eccentricity and semi-latus rectum parameters. As the swirling parameter increases, the relative variations in the eccentricity increase, while the variations in the semi-latus rectum decrease rapidly. These trends of the changes related to the orbital eccentricity and the semi-latus rectum with the swirling parameter resemble those observed with the MOG parameter in the Scalar-Tensor-Vector-Gravity (STVG) theory, but with different rates of change. Furthermore, our results also reveal that effects of the background swirling parameter on the relative variations in the  eccentricity and the semi-latus rectum are distinctly different from those of the black hole spin parameter. These results provide deeper insights into the properties of EMRI gravitational waves and the background's swirling.

\end{abstract}

\maketitle
\newpage

\section{Introduction}

Gravitational wave astrophysics has entered  a completely new era due to the direct detections of hundreds of gravitational
wave events \cite{ref1,ref2,ref3}. Among various sources of gravitational
waves, EMRI systems \cite{ref13, ref16, ref17, ref18, ref19, ref20, ref24, ref25, ref26, ref27, ref35,ref39,ref40,ref41,ref43,ref44} are of
particular interest to future space-based gravitational
wave detectors \cite{Berry31,ref12, LISA31, Danzmann31,Hu31,Gong31,Tianqin31} because their gravitational
waves are in the lower frequency band \cite{ref4,ref5,ref6,ref7,ref11} which are difficult to be explored by the current ground-based detectors.
In the EMRI system,  a stellar-mass object (the secondary with mass $\mu$) slowly spirals inward the central supermassive black hole (the
primary with mass $M$) due to the gravitational emission, and undergoes numerous orbits before ultimate plunging. During its evolution,
the gravitational waveforms generated by EMRIs are closely related to the trajectory of the smaller object, which imprints characteristic information on the spacetime geometry around the supermassive black hole. This means that EMRI gravitational waves provide invaluable probes to study the properties
of supermassive black holes and to examine gravitational theories. EMRI gravitational waves have extensively applied to constrain the nature of Kerr spacetime \cite{bd23,bd24,bd25,bd25s,bd25s1,bd25s2,bd25s3,bd25s4,bd25s5,bd25s6,bd25s7,bd25s8,bd25s9}, and to test theories of gravity include Einstein's general relativity, the Brans-Dicke theory \cite{bd26,bd27,bd27a}, and dynamical Chern-Simons gravity \cite{cs28,cs29,cs30}, quantum gravity \cite{cqg31, cqg32,cqg33} as well as Einstein-Cartan theory \cite{EC01,EC02,EC03,EC04,EC05,EC06}. Recently, EMRI gravitational waves have also served as a power tool to detect dark matter around supermassive black holes \cite{dm17,dm18,dm19} and to probe scalar charges induced by the dipole radiation \cite{scalar31,scalar32,scalar33,scalar34,scalar35,scalar36,scalar37,scalar38,scalar39,scalar40}. EMRI gravitational waves can be used as standard sirens to investigate the expansion history of the universe, and further to constrain cosmological parameters \cite{ref15}.

The Kerr black hole is an important vacuum solution of stationary and  axisymmetric black hole in  general relativity. Therefore, it is significant to study gravitational waves in the rotating vacuum black hole spacetimes deviated from the usual Kerr one because it provides some extra information on spacetime and gravity.
Recently,  another interesting and rotating vacuum solution of Einstein field equations was obtained by Astorino \textit{et al} \cite{ref14} through exploiting the Ernst formalism \cite{ref30,ref31} in combination with the Ehlers transformation by starting from a Kerr seed spacetime. This black hole  is characterized by three parameters: black hole mass and spin parameters, as well as the swirling parameter.  The swirling parameter describes
the  background's rotation, which can be interpreted as a gravitational whirlpool. The spin-spin interaction between the black hole and the background frame dragging results in the breaking of symmetry of spacetime geometry due to an additional force acting on the axis. Especially,
the swirling-Kerr black hole is co-rotating with the swirling background in one hemisphere
and counter-rotating in the other since the black hole spin does not affect its sign with respect to the equatorial
plane. The breaking of symmetry leads to the non-removable conical singularities along the symmetry axe, which means that
there exists a cosmic string or a strut to compensate the force effect induced by the spin-spin interaction.
The swirling-Kerr black hole solution possesses two horizon surfaces located at $r_{\pm}=M\pm \sqrt{M^2-a^2}$, which is independent of the swirling parameter and equivalent to the Kerr case. However, that the background swirling  deforms the horizon geometry \cite{ref14}.
In this swirling-Kerr black hole spacetime, there is no closed time-like curves which means that it is less exotic than the G\"{o}del or the
Taub-NUT spacetime \cite{ref14}. The geodesic in
the swirling universe is found to no longer be decoupled because the spacetime is no longer of Petrov type $D$, which could yield chaotic motion of particles in this spacetime \cite{chaos31,Geode}. Moreover, the swirling parameter is found to drive the light rings outside the equatorial plane, which results in that the contour of the shadow becomes a tilted oblate shape \cite{shadows1}. Recent studies  analyzed the geometrically thick
equilibrium tori orbiting a Schwarzschild black hole with the swirling parameter \cite{disk1} and a swirling-Kerr black hole \cite{disk2}
respectively, and explored new effects arising from background swirling on the equilibrium tori around the black hole. The studies of the swirling black holes have been generalized to the electromagnetic case \cite{ref28,ref29}. These studies shed new light on understanding black holes with background's swirling.

The main motivation of this paper is to study the gravitational waves from equatorially eccentric EMRIs around a swirling-Kerr black hole and to probe what new effects arise from the background swirling and differ from those of Kerr black hole. For EMRIs, the direct numerical relativity simulations  are computationally prohibitive due to the vast separation of timescales and spatial scales. The numerical Kludge approach addresses this by hybridizing multiple approximate theories to generate sufficiently accurate gravitational waveforms templates at a fraction of the computational cost. Therefore, we here employ the numerical kludge (NK) scheme  \cite{ref9,ref36} to simulate gravitational waveforms generated by this EMRI system and discuss  how the background swirling
influence these waveforms. We further also analyzed the gravitational waveform confusion problem.

The plan of this paper is as follows: In the Sec.II, we briefly review the swirling-Kerr black hole solution \cite{ref14}  and the equatorial EMRI systems. In Sec.III, we present properties of EMRI gravitational waves in the swirling background and analyze effects of the swirling parameter as well as the black hole spin parameter. Finally, we present a summary.

\section{Equatorial EMRI system within the framework of swirling Kerr spacetimes}

Lets us first briefly review the Kerr black hole solution embedded in a rotating  background. It is a stationary, axially symmetric and non-asymptotically flat black hole solution of the vacuum Einstein equations, which can be obtained through an Ehlers transformation on the Kerr solution as a seed \cite{ref14}. The swirling Kerr black hole possesses the swirling parameter and the spin parameter, and its metric form in the Boyer-Lindquist
coordinates can be written as
\begin{equation}
    \begin{aligned}
        ds^2=F(d\phi-\omega dt)^2+F^{-1}\bigg[-\rho^2dt^2+\Sigma \sin^{2}\theta(\frac{dr^2}{\Delta}+d\theta^2)\bigg],
    \end{aligned}
\end{equation}
where functions $F$ and $\omega$ can be expressed as a finite power series of the swirling parameter $j$
\begin{equation}
        F^{-1}=\chi_{(0)}+j\chi_{(1)}+j^2\chi_{(2)},\quad\quad\quad \omega=\omega_{(0)}+j\omega_{(1)}+j^2\omega_{(2)}.
\end{equation}
The expansion coefficients $\chi_{(i)}$ and $\omega_{(i)}$ are
\begin{equation}
\chi_{(0)}=\frac{R^2}{\Sigma \sin^2\theta},\quad\quad\quad \chi_{(1)}=\frac{4aM\Xi\cos{\theta}}{\Sigma \sin^{2}\theta},\quad\quad\quad
\chi_{(2)}=\frac{4a^{2}M^{2}\Xi^{2}\cos^2{\theta}+\Sigma^{2}\sin^{4}{\theta}}{R^2\Sigma \sin^2{\theta}},
\end{equation}
and
\begin{eqnarray}
        \omega_{(0)}&=&\frac{2aMr}{-\Sigma},\quad\quad\quad\omega_{(1)}=\frac{4\cos{\theta}[-a\Omega(r-M)+Ma^4-r^4(r-2M)-\Delta a^2 r]}{-\Sigma},\nonumber\\
        \omega_{(2)}&=&-\frac{2M\{3ar^5-a^5(r+2M)+2a^3r^2(r+3M)-r^3(\cos^2{\theta}-6)\Omega+a^2[\cos^2{\theta}(3r-2M)-6(r-M)]\Omega\}}{\Sigma},\nonumber\\
\end{eqnarray}
where
\begin{eqnarray}
\Delta&=&r^2-2Mr+a^2,\quad \quad \quad \rho^2=\Delta\sin^2{\theta},\quad \quad \quad \Sigma=(r^2+a^2)^2-\Delta a^2\sin^2{\theta},  \nonumber \\
\Omega&=&\Delta a \cos^2{\theta},\quad \quad \quad  R^2=r^2+a^2\cos^2{\theta},\quad \quad \quad \Xi=r^2(\cos^2{\theta}-3)-a^2(1+\cos^2{\theta}).
\end{eqnarray}
Here $M$ and $a$ denote respectively the mass parameter of the black hole and the angular momentum per unit mass. It
reduces to the pure Kerr black hole as the swirling parameter $j = 0$ and to the Schwarzschild black hole in swirling universes as the black hole spin
$a =0$. The spin-spin interaction between the black hole and the background dragging
leads to a conical singularity along the symmetry axes \cite{ref14}, which deforms the horizon geometry and enhances the
symmetry breaking regarding the spacetime properties. The presence of $j$ yields that the north and south hemispheres of the swirling Kerr black hole rotate in opposite directions, which is different from that in the pure Kerr black hole case.

In the swirling Kerr black hole spacetime, the Lagrangian of a timelike particle moving along the geodesic is
\begin{eqnarray}
\mathcal{L}=\frac{1}{2}g_{\mu\nu}\dot{x}^{\mu}\dot{x}^{\nu}.
\end{eqnarray}
This Lagrangian is independent of the coordinates $t$ and $\phi$, which means that there are two conserved quantities for the timelike particle's motion along geodesics
\begin{equation}
E=- u_{t}=-g_{tt}u^t-g_{t\phi}u^\phi,\quad \quad \quad L_{z}=u_{\phi}=g_{t\phi}u^t+g_{\phi\phi}u^\phi,
\end{equation}
where
\begin{equation}
u^t=\frac{F(E+L_z\omega)}{\rho^2}, \quad \quad \quad
u^{\phi}=\frac{L_z}{F}-\frac{F\omega(E+L_z\omega)}{\rho^2}.
\end{equation}
Here $E$ and $L_{z}$ correspond to the energy and the angular momentum of a particle, respectively. The timelike condition $g_{\mu\nu}u^\mu u^\nu=-1$, leads to
a constraint condition for the particle's motion
\begin{equation}\label{Hcon}
	h=g_{t t} \dot{t}^{2}+g_{r r} \dot{r}^{2}+g_{\theta \theta} \dot{\theta}^{2}+g_{\varphi\varphi} \dot{\varphi}^{2}+2 g_{t \varphi} \dot{t} \dot{\varphi}+1=0.
\end{equation}
Generally, this first-order differential equation is non-separable and there is no the Carter constant $Q$ for the particle's motion. Therefore, we consider only the case where the particle's geodesics are limited in the equatorial plane and probe the corresponding effects of the swirling parameter on gravitational waves from extreme mass ratio inspirals.

For a bounded Kepler geodesic in the equatorial plane, the orbit can be characterized by the eccentricity $e$ and the semi-major axis $p$ with the forms
\begin{equation}
    e=\frac{r_{a}-r_{p}}{r_{a}+r_{p}},\quad \quad \quad  p=\frac{2r_{a}r_{p}}{r_{a}+r_{p}}.
\end{equation}
Here $r_{a}$ and $r_{p}$ respectively  represent the apastron and periastro in the geodesic orbit. For the equatorial orbits, the eccentricity $e$ and the semi-major axis $p$ can be determined by particle's $E$ and $L_{z}$. To circumvent numerical difficulties frequently related to turning points, one can transition the integration variable $r$ to the angular
coordinate $\chi$ \cite{ref9,ref36}
 \begin{equation}
    r=\frac{p}{1+e\cos{\chi}}.
\end{equation}
As the parameter $\chi$ transforms between 0 and $2\pi$, the coordinate $r$ travels back and forth between the apostolic points $r_{a}$ and $r_{p}$. The effective potential $V_{r}$ for the geodesics in the equatorial plane is
\begin{equation}
    V_{r}=-1-\frac{L_{z}^2}{F}+\frac{F(E+L_{z}\omega)^{2}}{\rho^2}.
\end{equation}
With the conditions $V_{r}(r_{a})$=0 and $V_{r}(r_{p})$=0, one can find that the orbital parameters $e$ and $p$ can be written as a function of $E$ and $L_{z}$, and vice versa. For an equatorial orbit, its orbital frequency is characterized by $\Omega_{r}$ and $\Omega_{\phi}$, i.e.,
\begin{equation}\label{orbitfr1}
\begin{aligned}
    \Omega_{r}=\frac{2\pi}{T_{r}},\quad \quad \quad \Omega_{\phi}=\frac{\Delta\phi}{T_{r}}.
\end{aligned}
\end{equation}
Here the radial period $T_{r}$  and the circumstellar shift $\Delta \phi$  can be represented by simple integral expressions \cite{ref19,ref20}
\begin{equation}
    \begin{aligned}
        &T_{r}=\int_{r_{b}}^{r_{a}} \frac{u^{t}}{u^{r}} dr=\int_{0}^{2\pi} \frac{dr}{d\chi} \frac{u^{t}}{u^{r}} d\chi,\\
        &\Delta\phi=\int_{0}^{T_{r}} \frac{u^{\phi}}{u^{t}} dt=\int_{0}^{2\pi} \frac{dr}{d\chi} \frac{u^{\phi}}{u^{r}} d\chi.
    \end{aligned}
\end{equation}
The number of cycles $\mathcal{N}$ is a well quantity illustrated the difference between Kerr and swirling-Kerr
orbits, which requires to accumulate difference $\pi/2$ in periastron shift
\begin{equation}
    \begin{aligned}
        \mathcal{N}=\dfrac{\pi/2}{\lvert \Delta\phi_{K}-\Delta\phi_{sK} \lvert},
    \end{aligned}
\end{equation}
where the indices $K$ and $sK$ denote the Kerr and swirling-Kerr cases, respectively.
\begin{figure}\centering
\includegraphics[width=1\linewidth,height=0.9\linewidth]{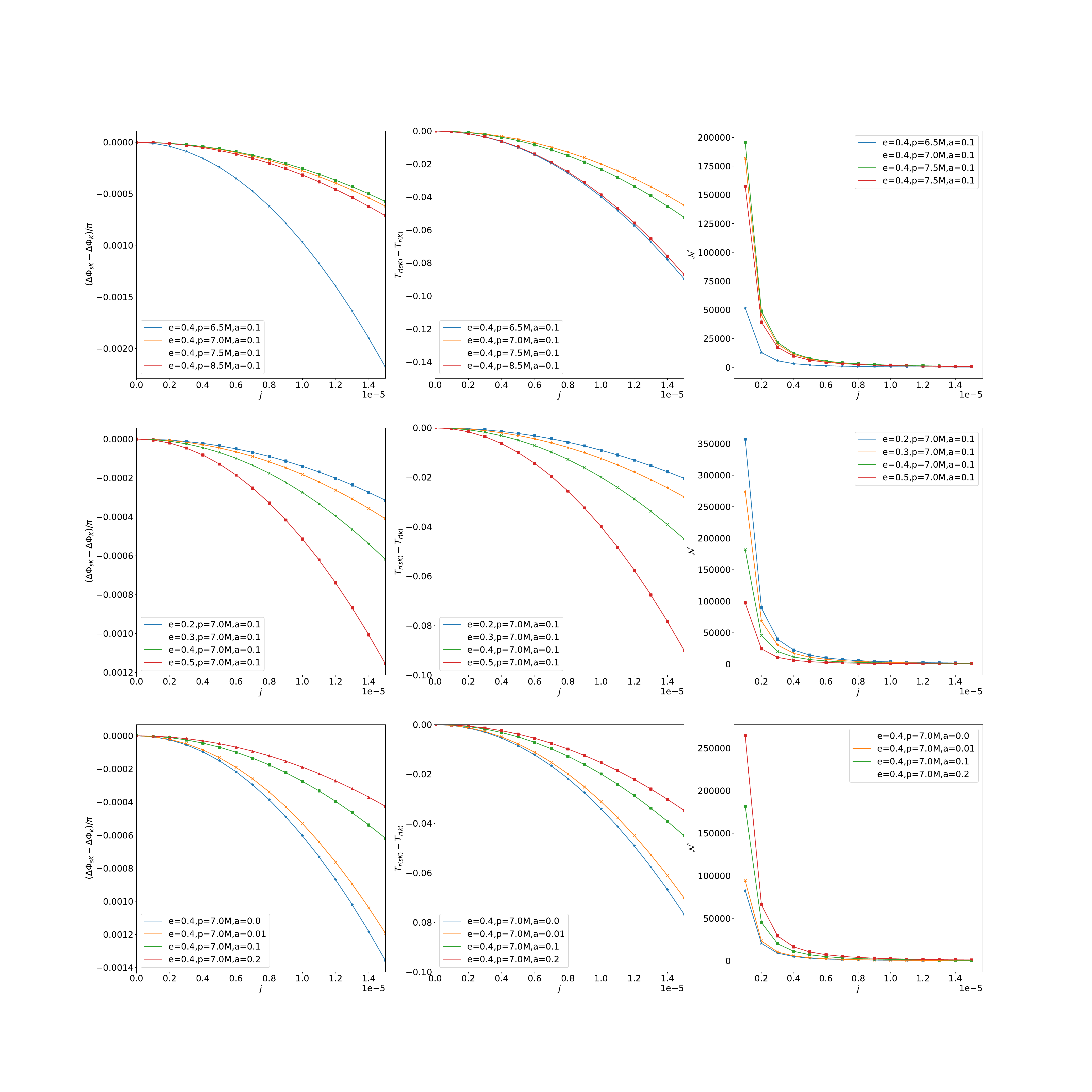}
\caption{Changes of the periastron shift difference $\Delta\phi_{sK}-\Delta\phi_{K}$ (left panel),  the radial period difference $T_{r(sK)}-T_{r(K)}$ (middle panel) and the number of cycles $N$ (right panel) with the swirling parameter $j$ for different $e$, $p$ and $a$.}
\label{fig1}
\end{figure}
The larger the number of cycles $\mathcal{N}$ means the smaller orbital deviations. Fig. \ref{fig1} shows effects of orbital parameters $e$, $p$ and black hole spin $a$ on the periastron shift difference $\Delta\phi_{sK}-\Delta\phi_{K}$ (left panel),  the radial period difference $T_{r(sK)}-T_{r(K)}$ (middle panel) and the number of cycles $\mathcal{N}$ for different swirling parameter $j$. Obviously, the absolute values of these orbital differences $|\Delta\phi_{sK}-\Delta\phi_{K}|$ and $|T_{r(sK)}-T_{r(K)}|$ increases with the swirling parameter $j$,  while the number of cycles $\mathcal{N}$ decreases.
The effects of the parameters $j$ is just opposite to  those of the black hole spin parameter $a$, which could provide a probe to identify the swirling of the universe background. Moreover, with increases of the parameter $p$, the quantities $|\Delta\phi_{sK}-\Delta\phi_{K}|$ and $|T_{r(sK)}-T_{r(K)}|$ first decreases and then increases, while the number of cycles $\mathcal{N}$ first increases and then decreases. With increase of the parameter $e$, the absolute values $|\Delta\phi_{sK}-\Delta\phi_{K}|$ and $|T_{r(sK)}-T_{r(K)}|$ increase and the number of cycles $\mathcal{N}$ decreases.

We are now in position to compute the gravitational waveforms caused by the particle's trajectory in the swirling-Kerr spacetime
with the kludge waveform generation method.  As in Ref. \cite{ref32}, adopting Cartesian coordinates
$x=r\sin{\theta}\cos{\phi}$, $y=r\sin{\theta} \sin{\phi}$, $z=r\cos{\theta}$, and employing the multipolar expansion of metric perturbations,
one can obtain the lowest-order term in gravitational waves emitted by an isolated system through the quadrupole formula
\begin{equation}
    \begin{aligned}
        \overline{h}^{jk}(t,x)=\frac{2}{R}[\Ddot{I}^{jk}(t^{'})]_{t^{'}=t-r},
    \end{aligned}
\end{equation}
where $R$ denotes the luminosity distance from the source to the observer and $I^{jk}(t')$ is the source's mass quadrupole moment with the form
\begin{equation}
    \begin{aligned}
        I^{jk}(t')=\int x'^{j}x'^{k}T^{00}(t',x') d^{3}x'.
    \end{aligned}
\end{equation}
Following the procedure proposed in Ref. \cite{ref9,ref36}, with the transverse-traceless
(TT) projection to the previously mentioned expressions and the
standard TT gauge,  the plus and cross components of the waveform observed at latitude $\Theta$ and azimuth $\Phi$ can be expressed as
\begin{eqnarray}
        h_{+}&=&h^{\Theta\Theta}-h^{\Phi\Phi}=\{\cos^2{\Theta}[h^{xx}\cos^2{\Phi}+h^{xy}\sin{2\Phi} h^{yy} \sin^2{\Phi}]+h^{zz}\sin^2{\Theta}-\sin{2\Theta}[h^{xz}\cos{\Phi}+h^{yz}\cos{\Phi}]\}\nonumber\\
        &-&[h^{xx}\sin^2{\Phi}-h^{xy}\sin{2\Phi}+h^{yy}\cos^2{\Phi}],\nonumber\\
h_{\times}&=&2h^{\Theta\Phi}=2\{\cos{\Theta}[-\frac{1}{2}h^{xx}\sin{2\Phi}+h^{xy}\cos{2\Phi}+\frac{1}{2}h^{yy}\sin{2\Phi}]
+\sin{\Theta}[h^{xz}\sin{\Phi}-h^{yz}\cos{\Phi}]\}.
\end{eqnarray}
To analyze the distinctions between waveforms in swirling Kerr spacetimes and those
predicted from general relativity, we can utilize the overlap function to assess differences \cite{ref20},
\begin{equation}
    \begin{aligned}
      \mathcal{F}(\Tilde{h}_{1},\Tilde{h}_{2})=\frac{(\Tilde{h}_{1}\mid \Tilde{h}_{2})}{\sqrt{(\Tilde{h}_{1}\mid \Tilde{h}_{1}) (\Tilde{h}_{2}\mid \Tilde{h}_{2})}},
    \end{aligned}\label{overlapfunction}
\end{equation}
where the quantity marked with ``tilde" $\tilde{h}$ denotes the frequency domain signal related to the time domain signal $h(t)$ of gravitational wave by the Fourier transforms. The inner product $(\Tilde{h}_{1},\Tilde{h}_{2})$ is defined by \cite{ref10,ref21,ref22,ref23,ref33,ref34}
\begin{equation}
    \begin{aligned}
        (\Tilde{h}_{1} \mid \Tilde{h}_{2})=2\int_{0}^{\infty} \frac{\Tilde{h}_{1}^{*}(f)\Tilde{h}_{2}(f)+\Tilde{h}_{1}(f) \Tilde{h}_{2}^{*}(f)}{S_{n}(f)} df,
    \end{aligned}
\end{equation}
where $S_{n}(f)$ is the signal-to-noise ratio of the detector.  The noise spectral density of LISA is \cite{ref22,ref23}
\begin{equation}
    \begin{aligned}
        S_{n}(f)=\frac{1}{L^2}\{S_{x}+[1+(\frac{0.4mHz}{f}^{2})\frac{4S_{a}}{(2\pi f)^{4}}]\}.
    \end{aligned}.
\end{equation}
The overlap function $\mathcal{F}(\Tilde{h}_{1},\Tilde{h}_{2})=1$ if the two waveforms are identical, while $\mathcal{F}(\Tilde{h}_{1},\Tilde{h}_{2})=0$  if they are totally uncorrelated, and $\mathcal{F}(\Tilde{h}_{1},\Tilde{h}_{2})=-1$ if they are perfectly anti-correlated \cite{ref24}. Therefore, the higher value of overlap function means that it is more difficult to distinguish these two waveform signals.

\section{Effects of the swirling parameter on EMRI gravitational waves}

Let us now to probe effects of the swirling parameter together with the black hole spin parameter on EMRI gravitational waves using the numerical Kludge method.
In our numerical analysis, the primary mass $M$ and the secondary mass as $\mu$ are respectively set as $M = 2\times10^5 M_{\odot}$ (solar mass) and $\mu = 2M_{\odot}$, which means the mass ratio $\eta = \mu/M = 10^{-5}$. The observed angular parameters are set to $\Theta=\pi/4$, $\Phi=0$, and the luminosity distance is $ D_L = 5Gpc$. To evaluate the effect of the rotation parameter $j$ on the EMRI gravitational waves, we focus only on orbital evolutions on the equatorial plane, which are determined by conservative dynamics without explicitly considering radiative reactions.

The initial values of $(E,L_{z},Q)$  for the equatorial geodesic orbits in the context of conservative dynamics are given by  \cite{ref36}
\begin{equation}
    \begin{aligned}
        V_r(E,L_{z},Q=0,r=r_{a})=0, \quad \quad  V_r(E,L_{z},Q=0,r=r_{p})=0, \quad \quad  Q=0.
    \end{aligned}
\end{equation}

\begin{figure}
	\begin{minipage}{0.49\linewidth}
		\centering
		\includegraphics[width=1\linewidth,height=0.35\linewidth]{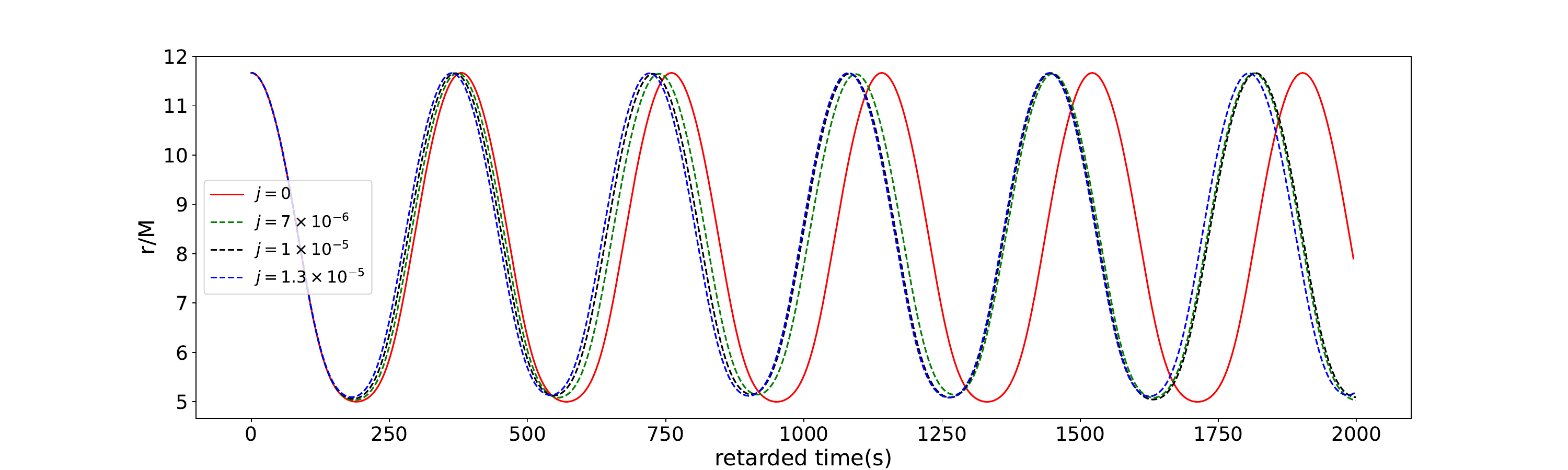}
	\end{minipage}
	\begin{minipage}{0.49\linewidth}
		\centering
		\includegraphics[width=1\linewidth,height=0.35\linewidth]{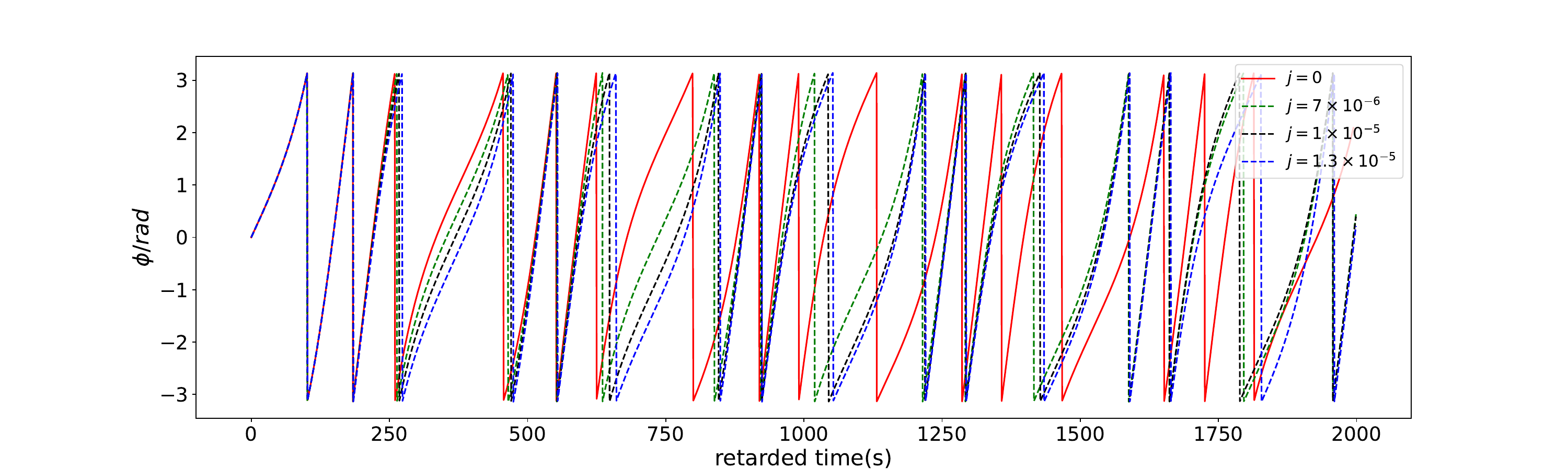}
     \end{minipage}
     \begin{minipage}{0.49\linewidth}
		\centering
		\includegraphics[width=1\linewidth,height=0.35\linewidth]{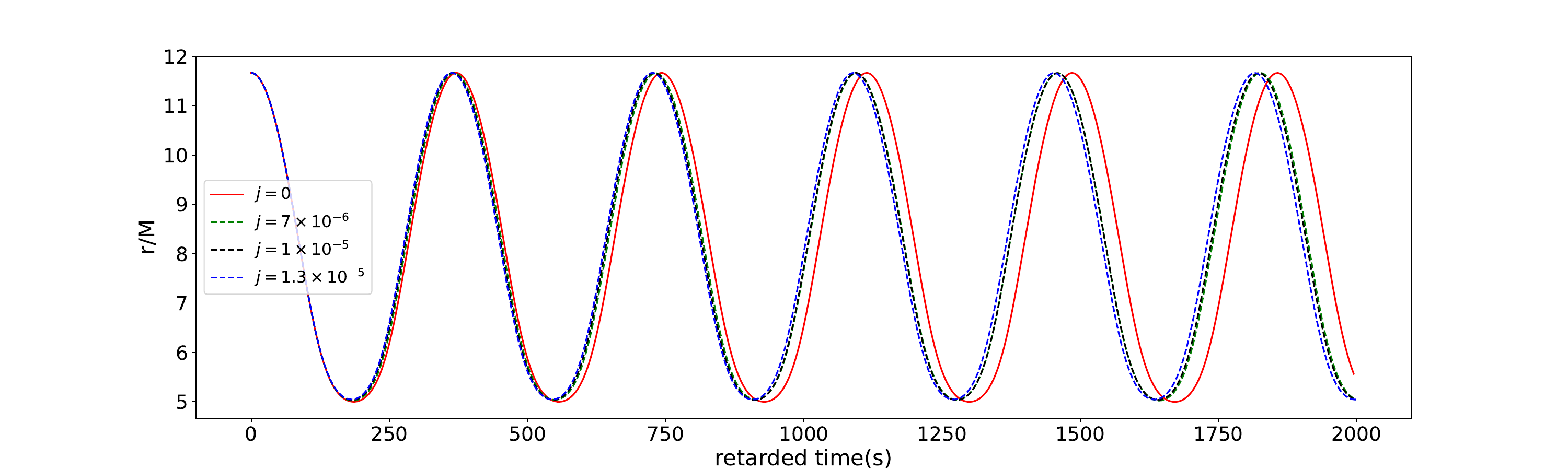}
     \end{minipage}
     \begin{minipage}{0.49\linewidth}
		\centering
		\includegraphics[width=1\linewidth,height=0.35\linewidth]{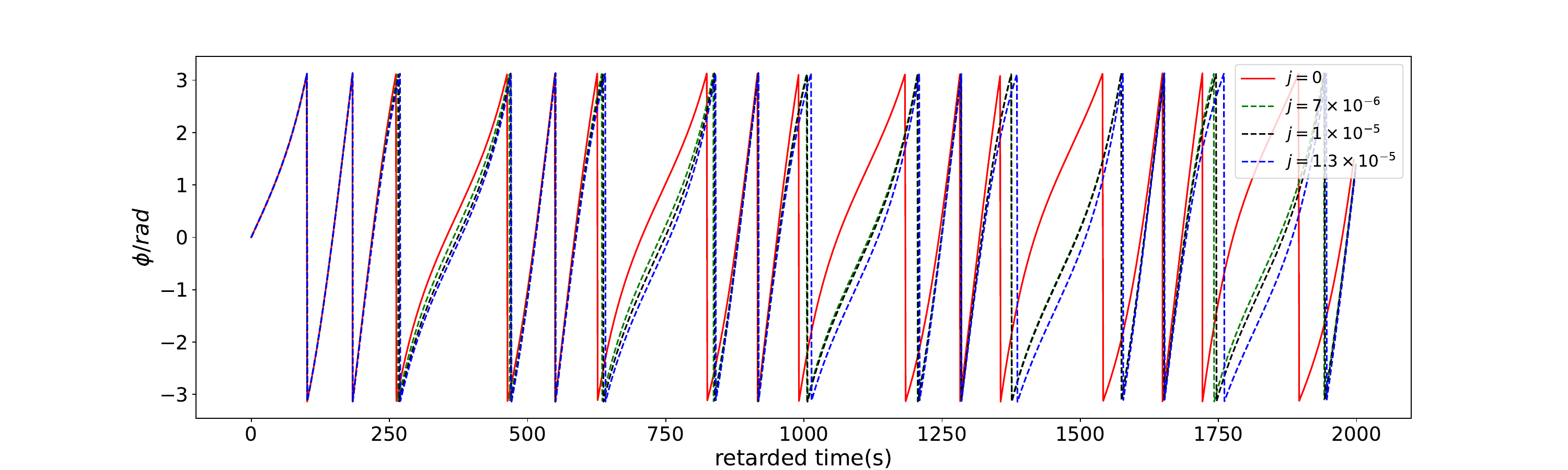}
     \end{minipage}
     \begin{minipage}{0.49\linewidth}
		\centering
		\includegraphics[width=1\linewidth,height=0.35\linewidth]{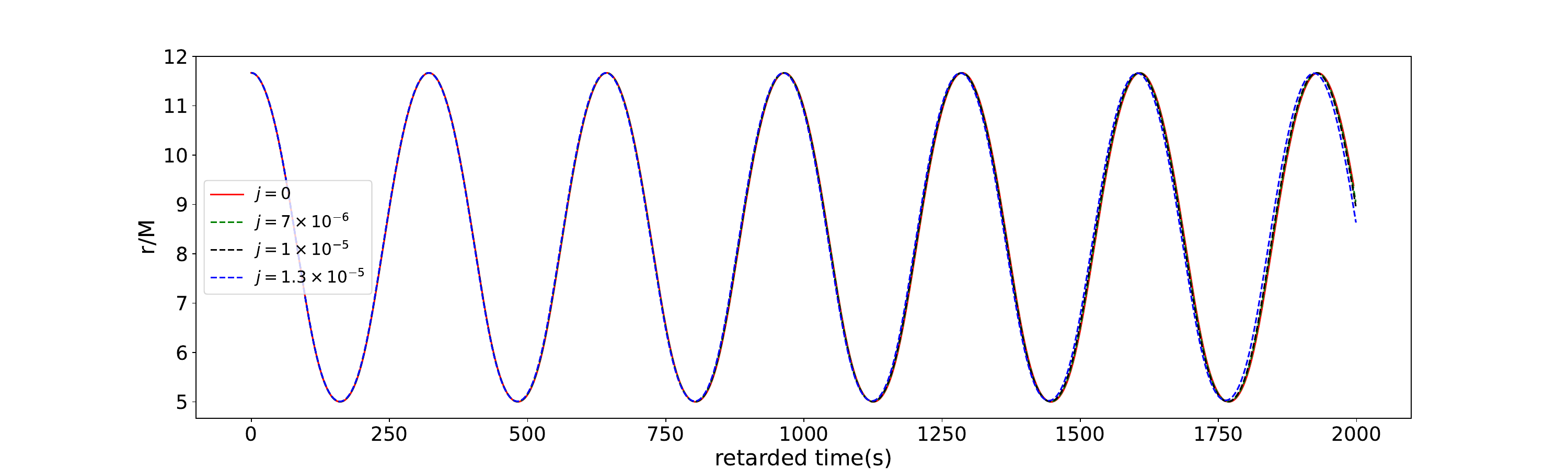}
     \end{minipage}
     \begin{minipage}{0.49\linewidth}
		\centering
		\includegraphics[width=1\linewidth,height=0.35\linewidth]{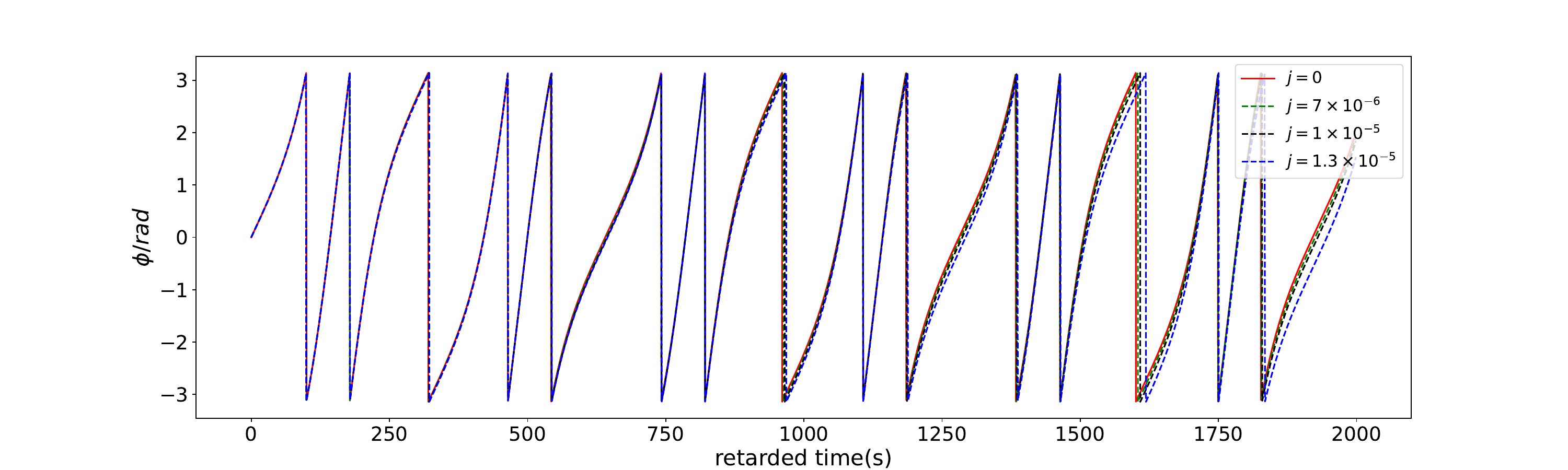}
     \end{minipage}
\caption{Time series of the orbital motion with the orbital parameters $e=0.4$ and $p=7.0M$ in the swirling Kerr spacetime for different $j$. The left panel and the right  panel respectively correspond to the motion in the $r$-direction and $\phi$-directions. The top, middle and bottom rows respectively denote $a=0.0$, $a=0.01$ and $a=0.1$.}
\label{fig2}
\end{figure}

\begin{figure}
	\centering
	\begin{minipage}{0.32\linewidth}
		\centering
		\includegraphics[width=0.9\linewidth,height=0.9\linewidth]{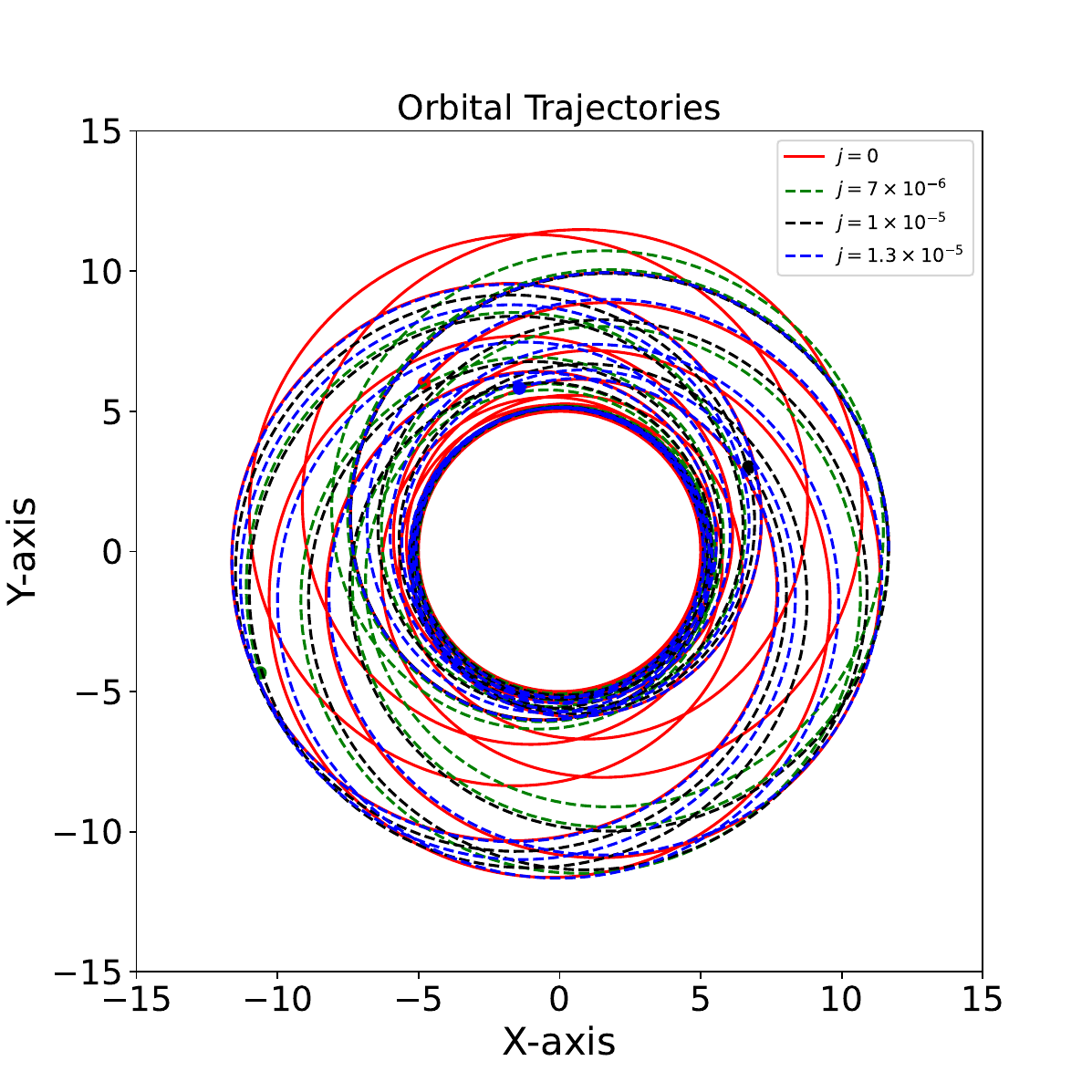}
	\end{minipage}
	\begin{minipage}{0.32\linewidth}
		\centering
		\includegraphics[width=0.9\linewidth,height=0.9\linewidth]{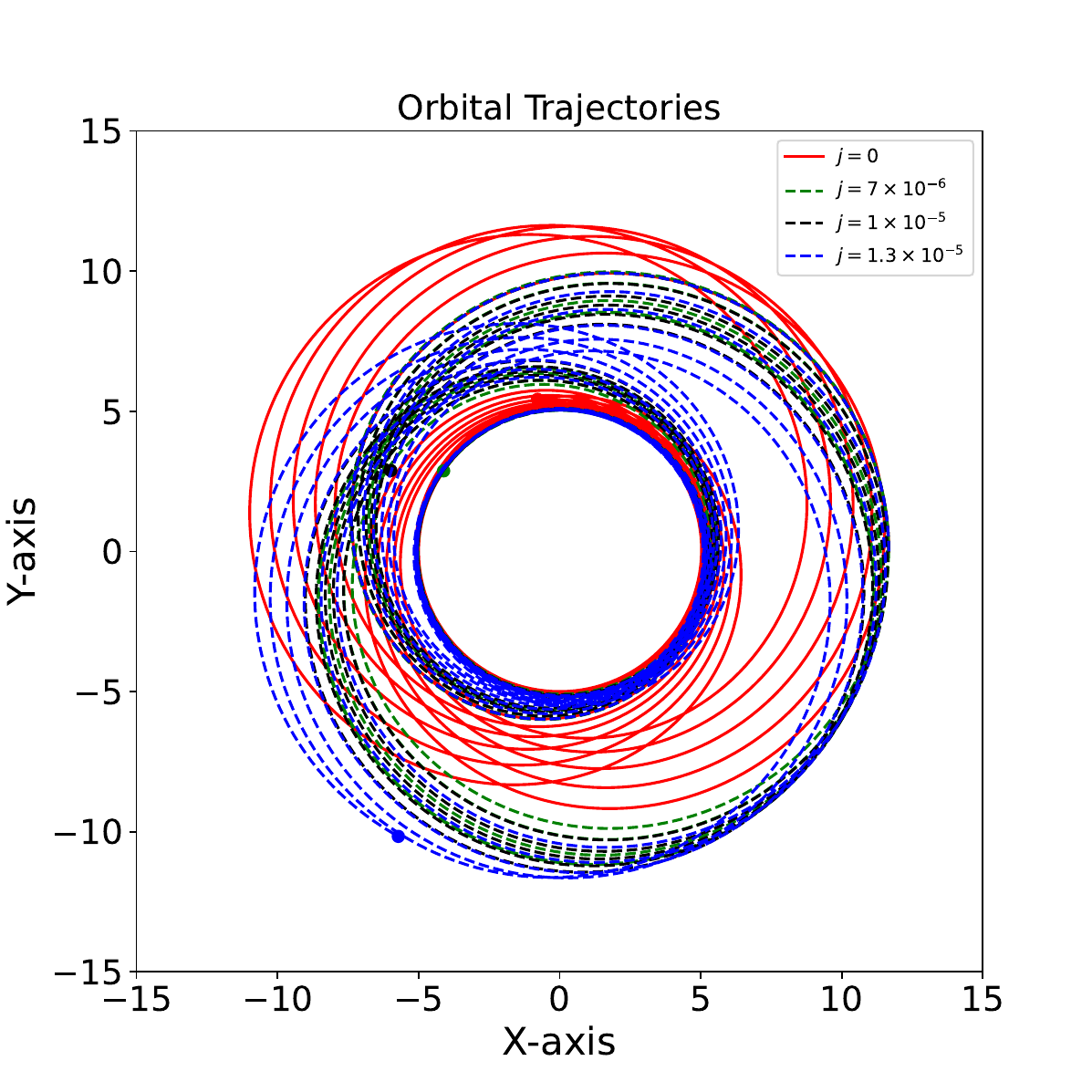}
	\end{minipage}
	\begin{minipage}{0.32\linewidth}
		\centering
		\includegraphics[width=0.9\linewidth,height=0.9\linewidth]{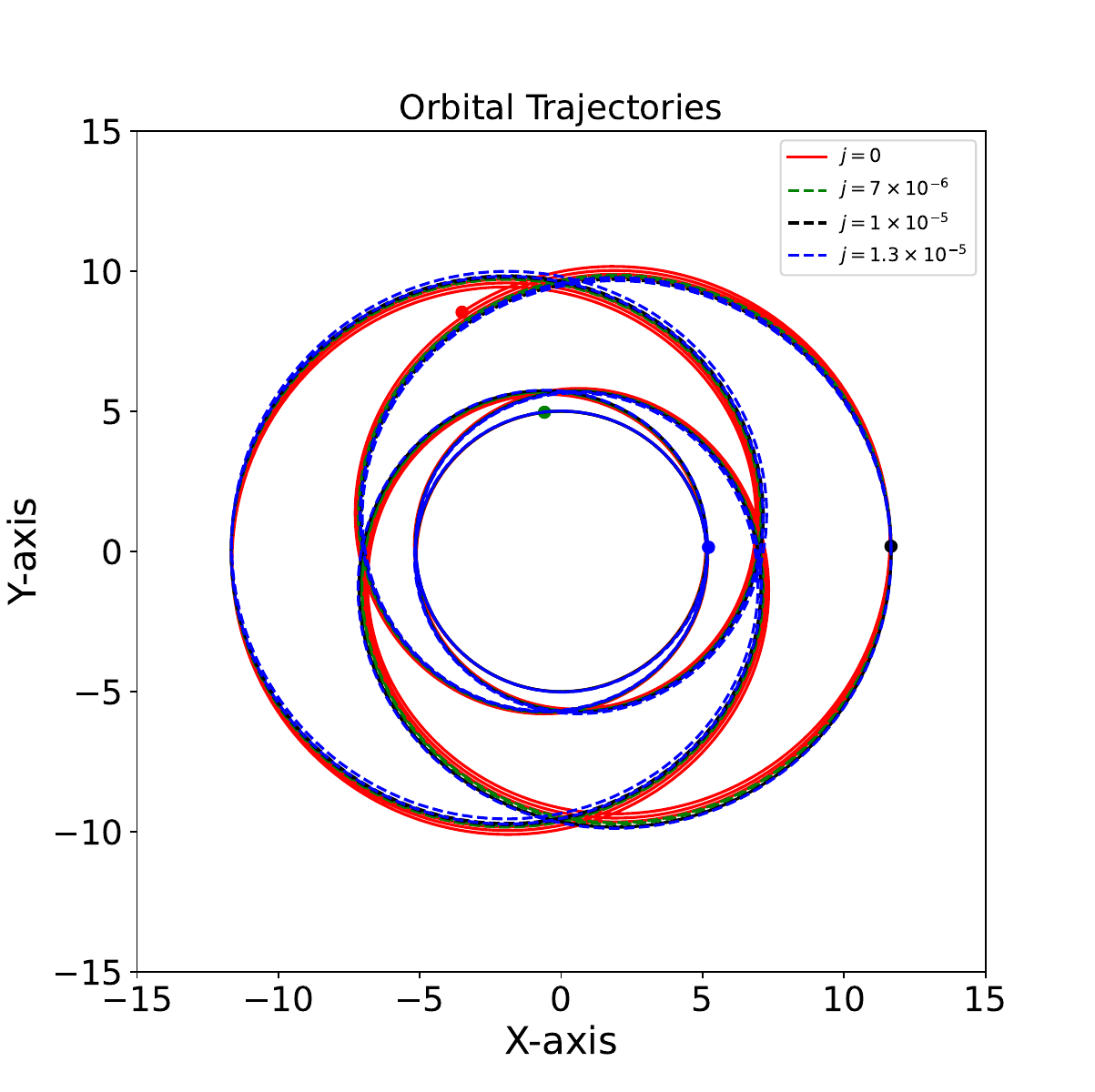}
	\end{minipage}
		\caption{Trajectories with the orbital parameters $e=0.4$ and $p=7.0M$ in the swirling Kerr spacetime for different swirling parameters $j$. The left, middle and right  panels respectively correspond to cases $a=0$, $a=0.01$ and $a=0.1$.}
        \label{fig3}
\end{figure}

Taking the same initial positions and orbital parameters $e$ and $p$, in Fig. \ref{fig2}, we present the time series of equatorially eccentric motion in the swirling Kerr black hole spacetime for different swirling parameter $j$ over the first 2000 seconds.
For fixed spin parameter $a$, we find that $j$ significantly influences the orbit periods and phase. As $j$ increases,the periods in the $r$ direction is decrease and in the $\phi$ direction increases. This observation is consistent with the results shown in Fig. \ref{fig1}.
With increase of the spin parameter $a$, the influence of the swirling parameter $j$ on the orbit decreases. As $a$ increases up to $a=0.1$, the difference arising from $j$ is almost indistinguishable in both the $r$ direction and $\phi$ direction. This means that in the rapidly rotating black hole case the effects of $j$ can be negligible. This is rational because in this case the frame dragging of the black hole  dominates over that of the background's swirling.
These properties of the spacetime parameters on the geodesic orbital trajectories are also shown in the Fig. \ref{fig3}.

\begin{figure}
	\centering
	\begin{minipage}{1\linewidth}
		\centering
		\includegraphics[width=1\linewidth,height=0.18\linewidth]{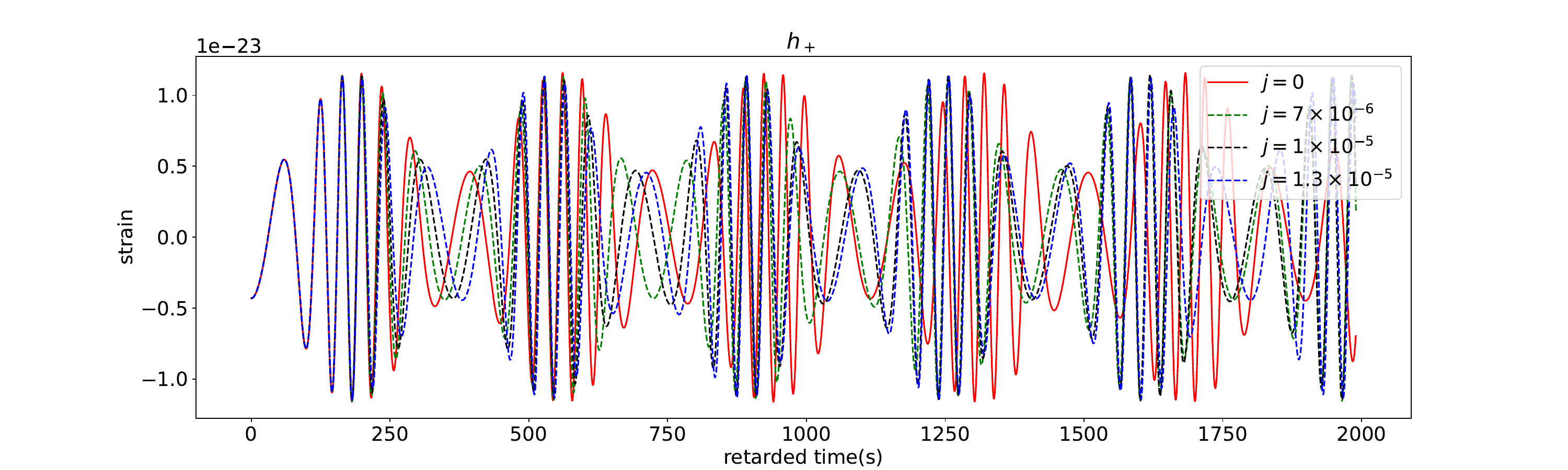}
	\end{minipage}
	\begin{minipage}{1\linewidth}
		\centering
		\includegraphics[width=1\linewidth,height=0.18\linewidth]{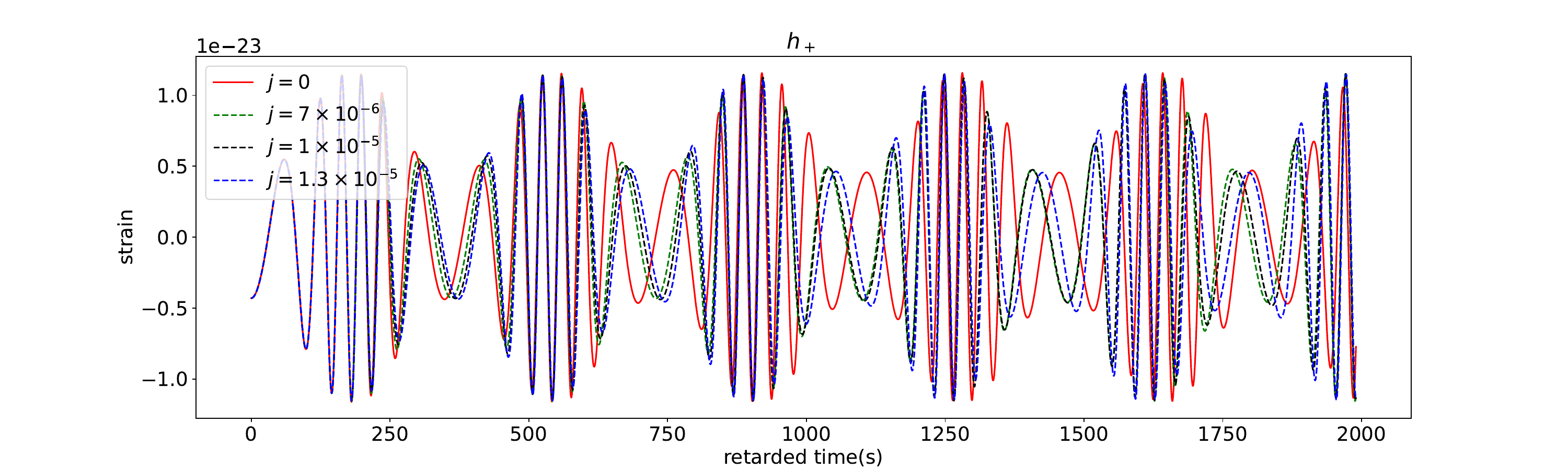}
	\end{minipage}
	\begin{minipage}{1\linewidth}
		\centering
		\includegraphics[width=1\linewidth,height=0.18\linewidth]{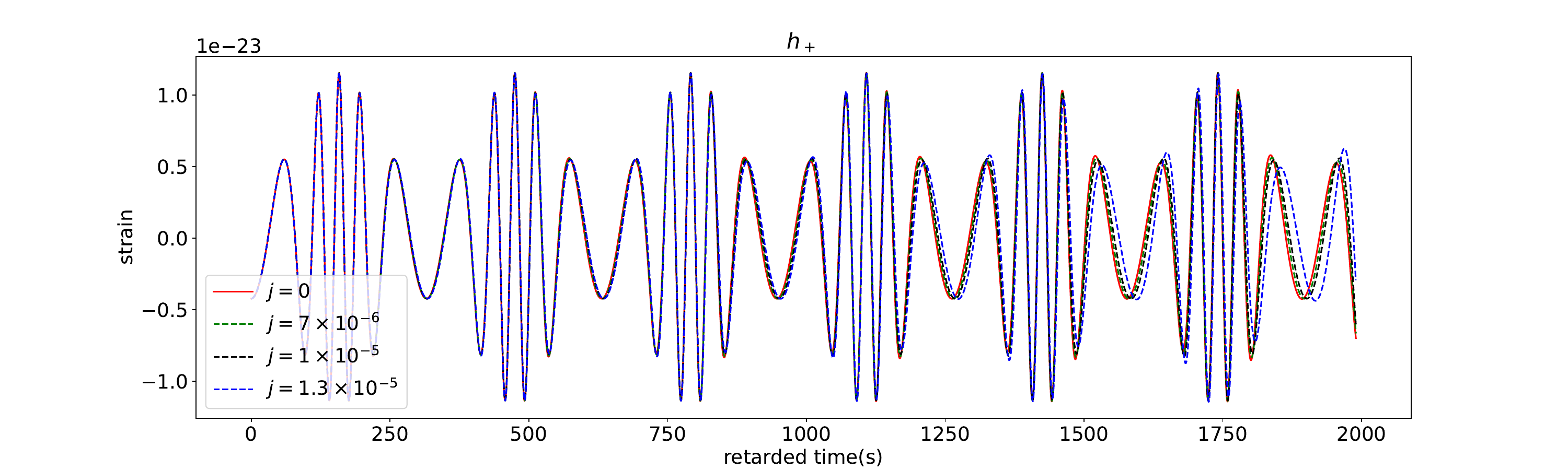}
	\end{minipage}
    	\caption{The plus component of EMRI waveforms for different swirling parameter $j$. The top, middle and bottom rows respectively correspond to the cases with centre black hole's spin $a=0.0$, $a=0.01$, $a=0.1$. Here, we set the orbital parameters to $e=0.4$ and $p=7.0M$.}
        \label{fig4}
\end{figure}

\begin{figure}
	\centering
	\begin{minipage}{1\linewidth}
		\centering
		\includegraphics[width=1\linewidth,height=0.18\linewidth]{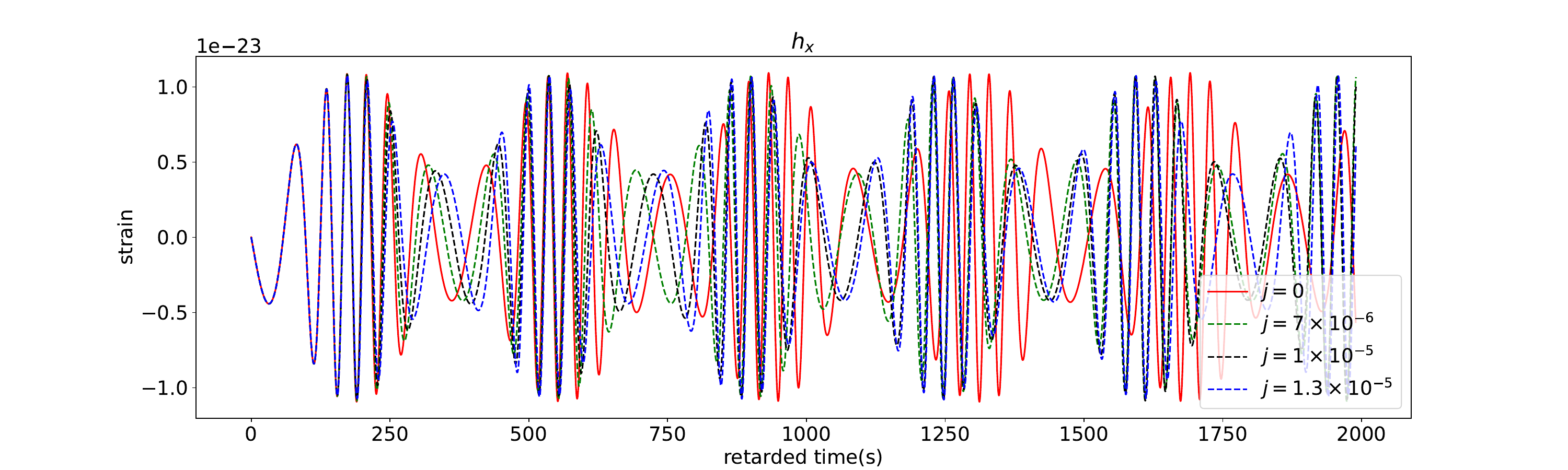}
	\end{minipage}
	\begin{minipage}{1\linewidth}
		\centering
		\includegraphics[width=1\linewidth,height=0.18\linewidth]{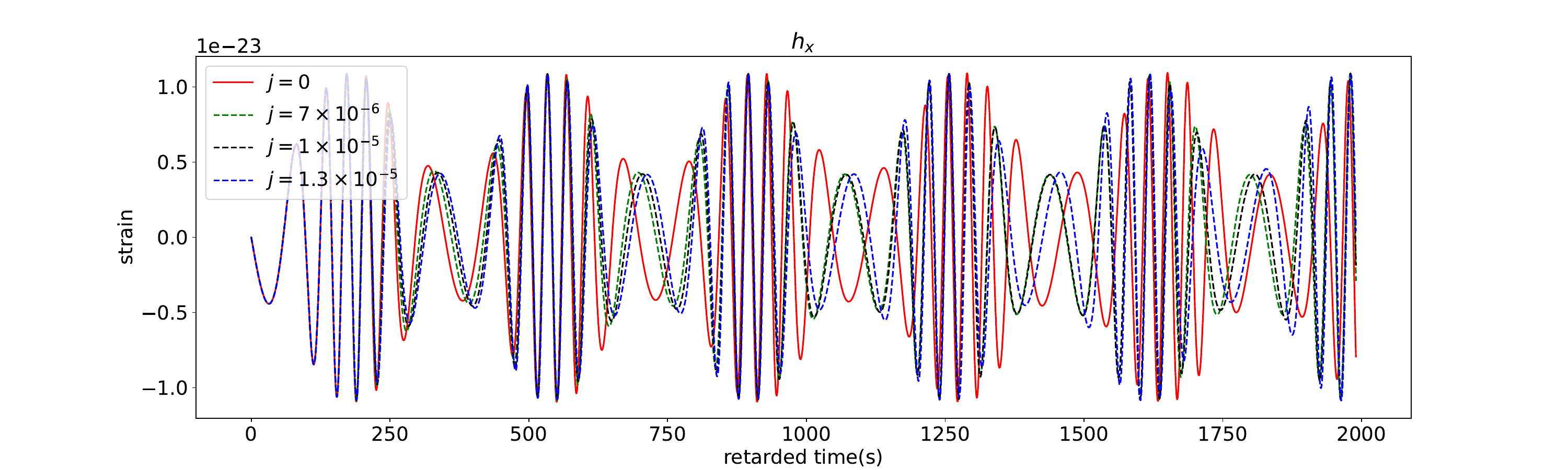}
	\end{minipage}
	\begin{minipage}{1\linewidth}
		\centering
		\includegraphics[width=1\linewidth,height=0.18\linewidth]{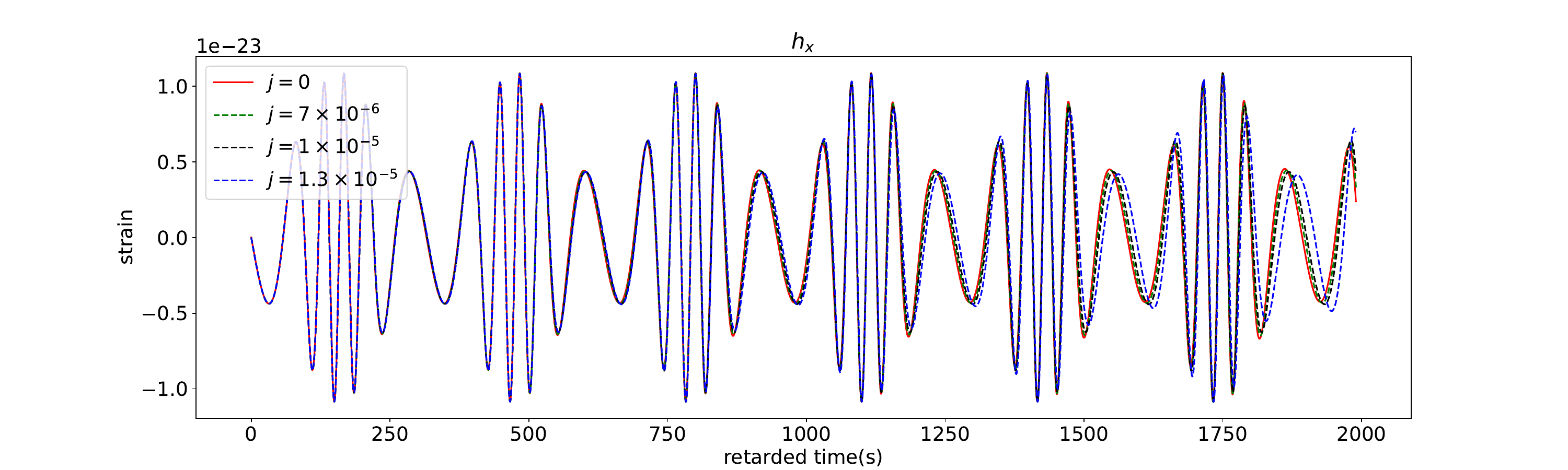}
	\end{minipage}
    	\caption{The cross component of EMRI waveforms for different swirling parameter $j$. The top, middle and bottom rows respectively correspond to the cases with centre black hole's spin $a=0.0$, $a=0.01$, $a=0.1$. Here, we set the orbital parameters to $e=0.4$ and $p=7.0M$.}
        \label{fig5}
\end{figure}

Based on the geodesic orbits given above, we can generate gravitational waveforms for plus and cross modes by using of the numerical Kludge method \cite{ref9,ref36}, which are respectively shown in Figs. \ref{fig4} and \ref{fig5}.
As $a=0$ and $a=0.01$, we find that the parameter $j$ has almost no effect on the gravitational waveforms during the initial 200 seconds, while the impact of the parameter $j$ becomes visually apparent after several orbital cycles and the parameter $j$ causes a phase delay of gravitational waveforms. As $a=0.1$, the gravitational waveform for different $j$ is visually almost identical in the initial 500 seconds and distinct difference arising from $j$ appear only after 1800 seconds. This means that the spin parameter $a$ suppresses the effects of the swirling parameter $j$ on the EMRI gravitational waves. Therefore, we can easily extract the information of the swirling parameter $j$ from the EMRI gravitational waves in the slowly rotating black hole case. For the case with higher spin parameter, in order to distinguish between the Kerr orbital waveform and the swirling-Kerr orbital waveform, one can need the longer retarded time for the EMRI gravitational wave to accumulate effects from the swirling parameter $j$.

\begin{figure}
	\centering
	\begin{minipage}{0.75\linewidth}
		\centering
		\includegraphics[width=0.8\linewidth,height=0.25\linewidth]{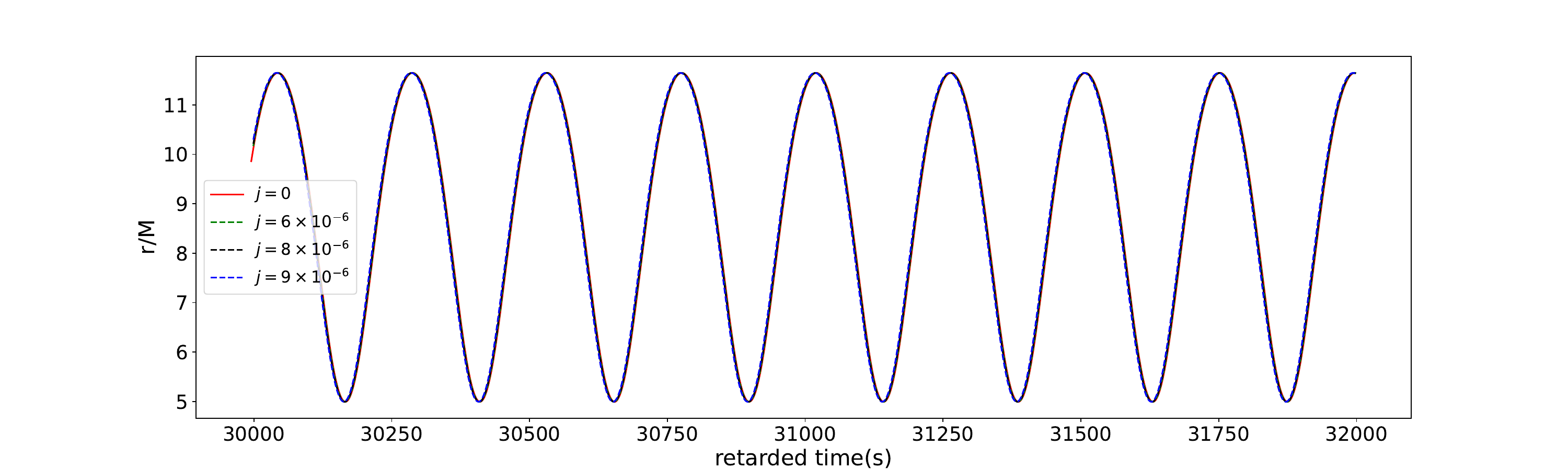}
	\end{minipage}
	\\
	\begin{minipage}{0.75\linewidth}
		\centering
		\includegraphics[width=0.8\linewidth,height=0.25\linewidth]{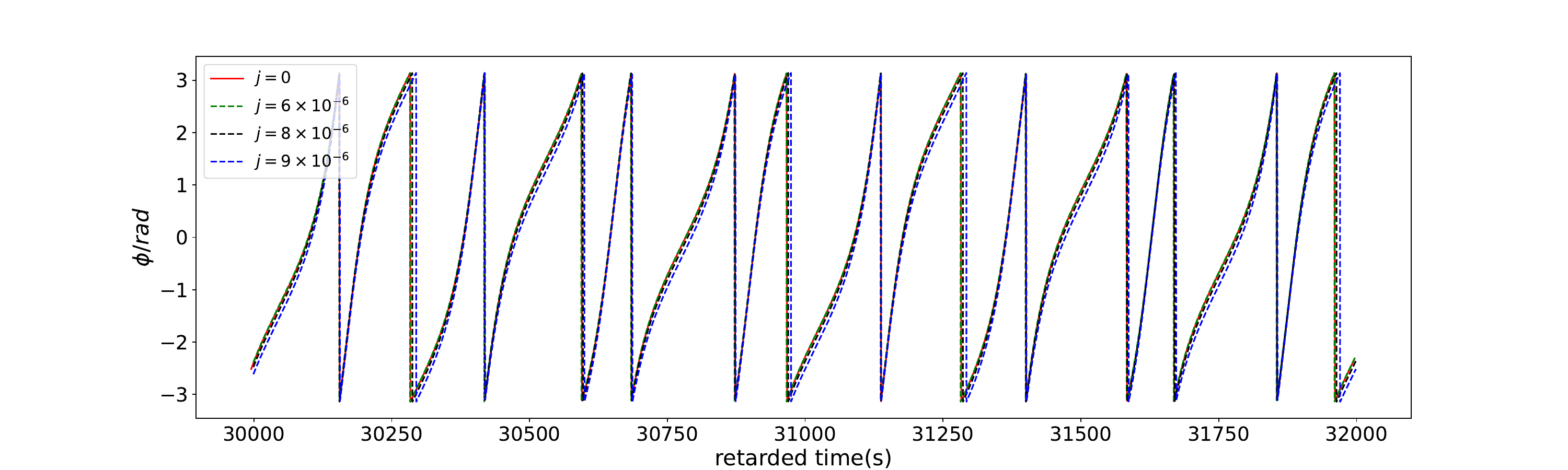}
	\end{minipage}
		\caption{Time series in the $r$ and $\phi$ directions under swirling-Kerr spacetime and Kerr spacetime with the same parameter set \{$e,p$,$a$\}=\{$0.4,7.0,0.5$\} for different swirling parameter $j$.}
        \label{fig6}
\end{figure}

As in other cases, the confusion problem is encountered when one identifies the signal of the swirling parameter from the EMRI gravitational wave data observed by real detectors. The main reason is that there exists the possibility of significant overlap between the EMRI gravitational waveforms
of Kerr and swirling-Kerr spacetimes  as they share the same orbital frequency. The significant overlap makes it difficult to effectively distinguish between the waveforms in Kerr and swirling-Kerr cases. As in the previous discussion, we noted that the lager spin parameter leads to confusion problem where the effects of the swirling parameter $j$ are heavily suppressed. In Fig.\ref{fig6}, we present the orbital deviation  between Kerr and swirling-Kerr spacetimes for the parameter set \{$e$, $p$, $a$\}=\{0.4,\;7.0,\;0.5\} for different $j$ and find that the two orbits are almost indistinguishable in this case.
\begin{figure}
    \includegraphics[width=7cm]{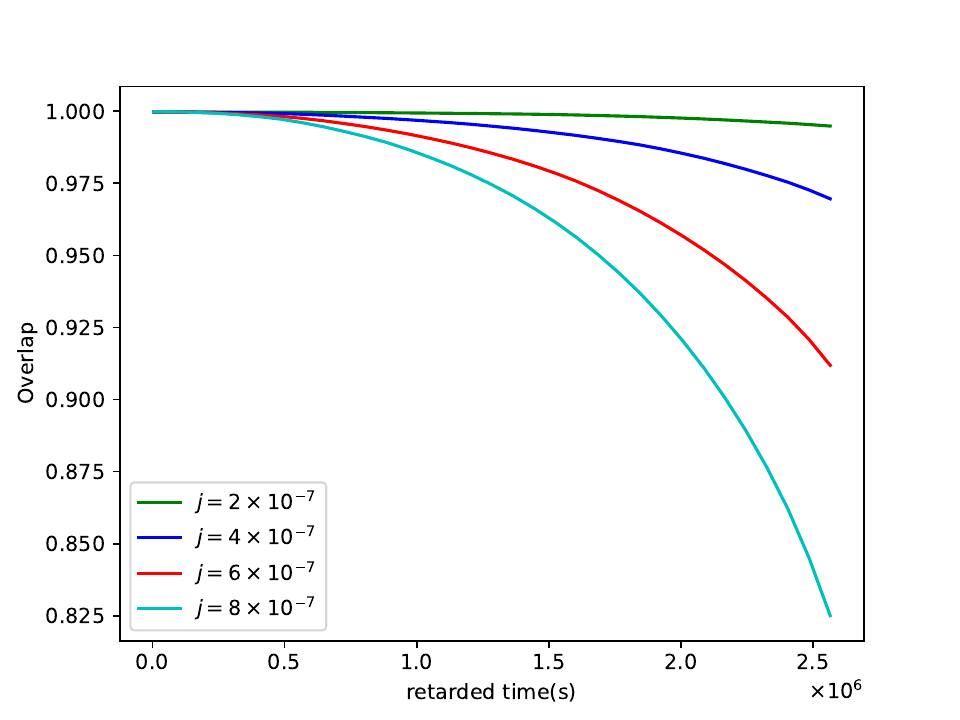}\includegraphics[width=7cm]{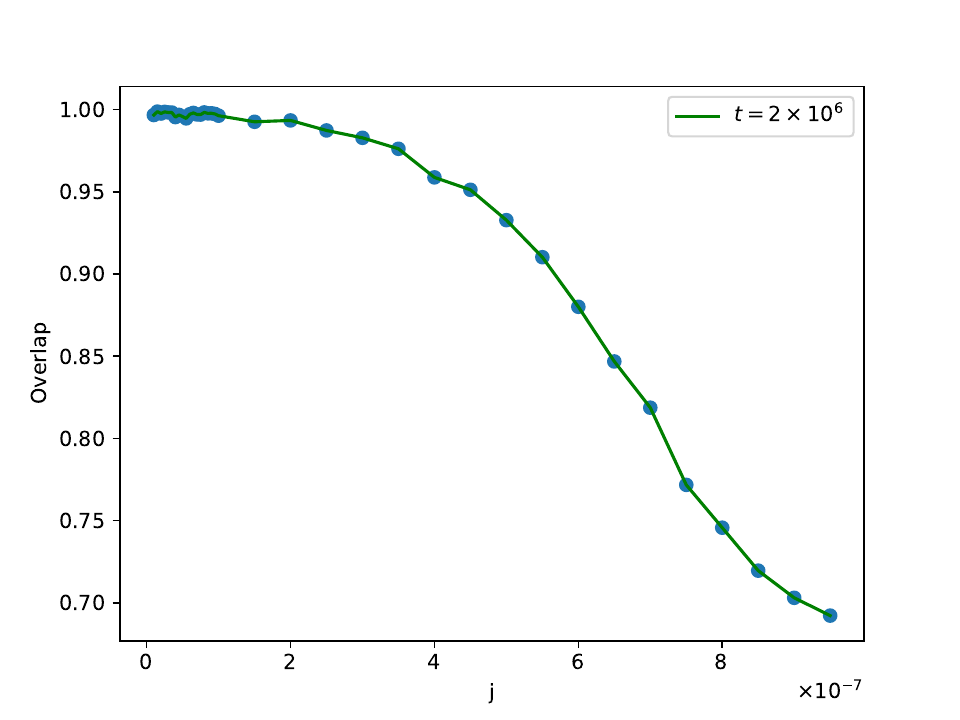}
	\caption{The overlap between Kerr and swirling-Kerr spacetimes with retarded time $t$ for different $j$ (left) and with $j$ for the fixed  signal length $t=2\times 10^{6}s$ (right). Here, the parameter set is \{$e,p$,$a$\}=\{$0.4,7.0,0.5$\}. }
        \label{fig7}
\end{figure}
In Fig. \ref{fig7}, we also present the overlap function (\ref{overlapfunction}) with the retarded time $t$ for different $j$ (left) and with $j$ for the fixed  signal length $t=2\times 10^{6}s$ (right). We find that the overlap value for $j=9 \times 10^{-7}$ is still very high ($\sim 0.8$) even if the retarded time $t=2 \times 10^6s$.  This
severely hinders our ability to detect the swirling parameter by analyzing EMRI gravitational waveforms. Moreover, Fig. \ref{fig7} also shows that the overlap function for the fixed  signal length also almost decreases with the swirling parameter $j$, which is rational  because the larger swirling parameter inevitably leads to greater differences between gravitational waveforms in Kerr and swirling-Kerr spacetimes.

\begin{figure}
	\centering
	\begin{minipage}{0.8\linewidth}
		\centering \includegraphics[width=0.33\linewidth,height=0.3\linewidth]{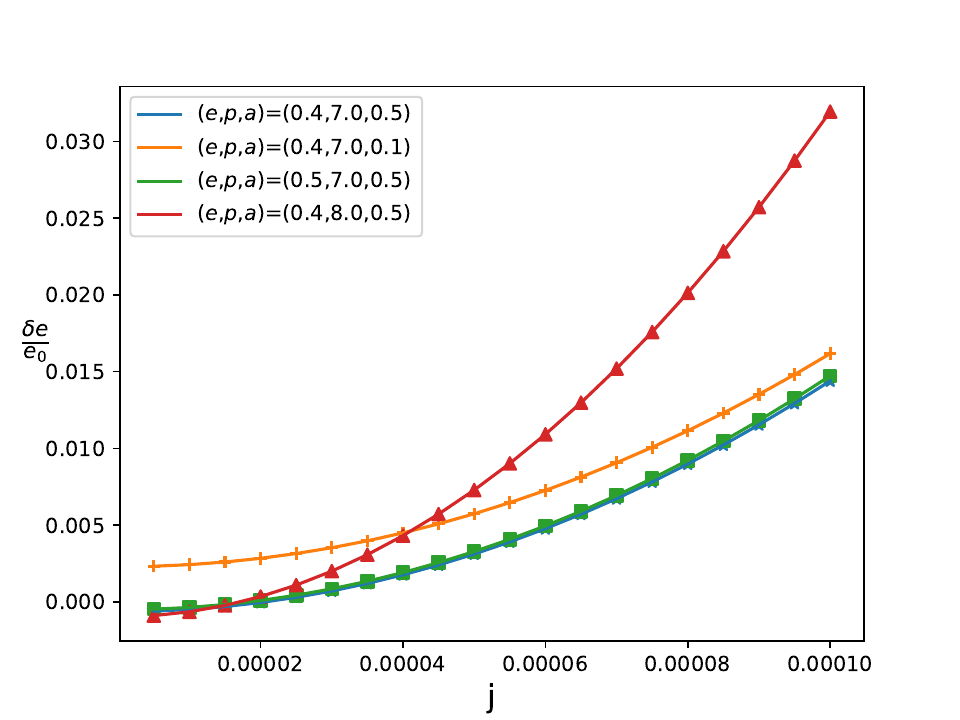}\includegraphics[width=0.332\linewidth,height=0.3\linewidth]{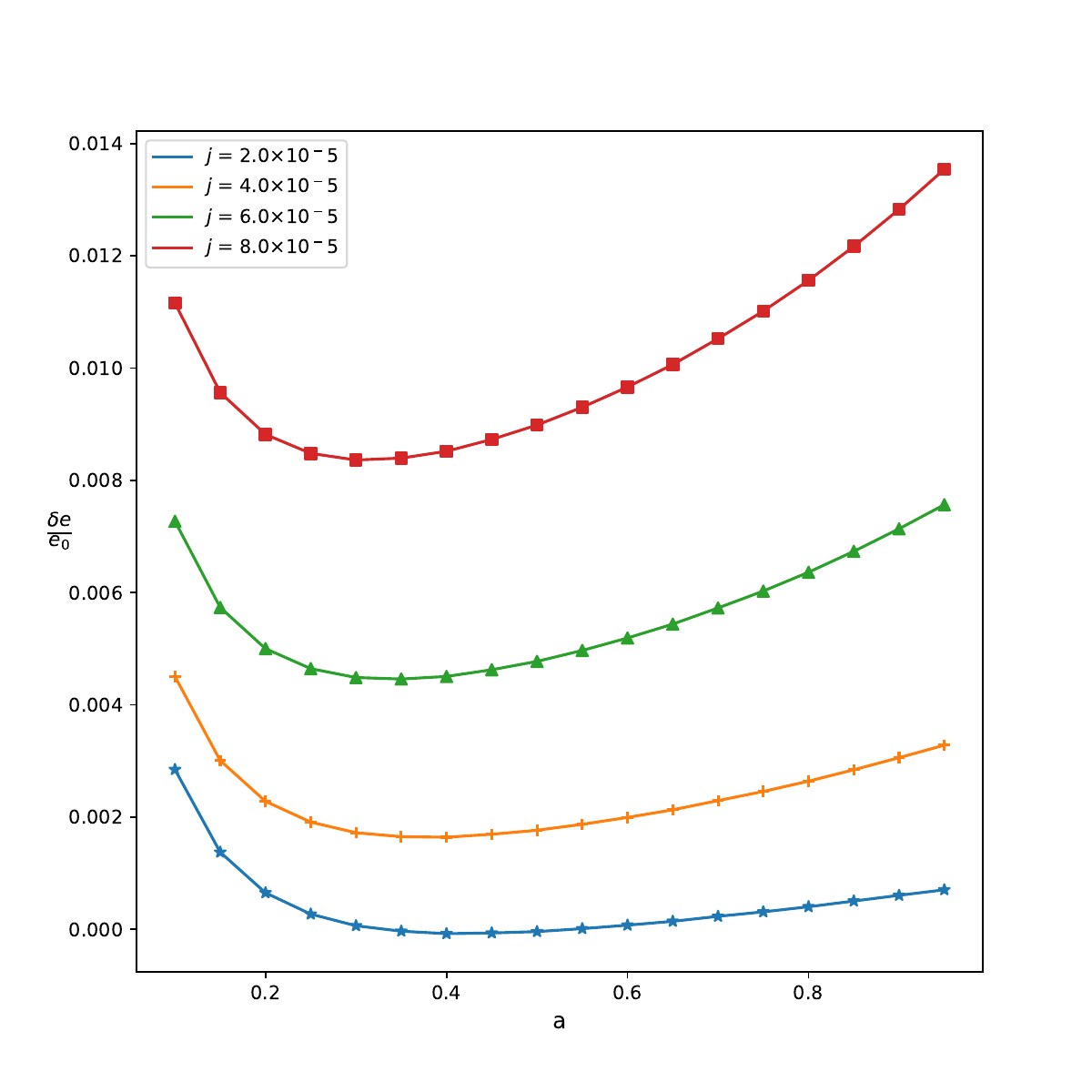}
\includegraphics[width=0.33\linewidth,height=0.3\linewidth]{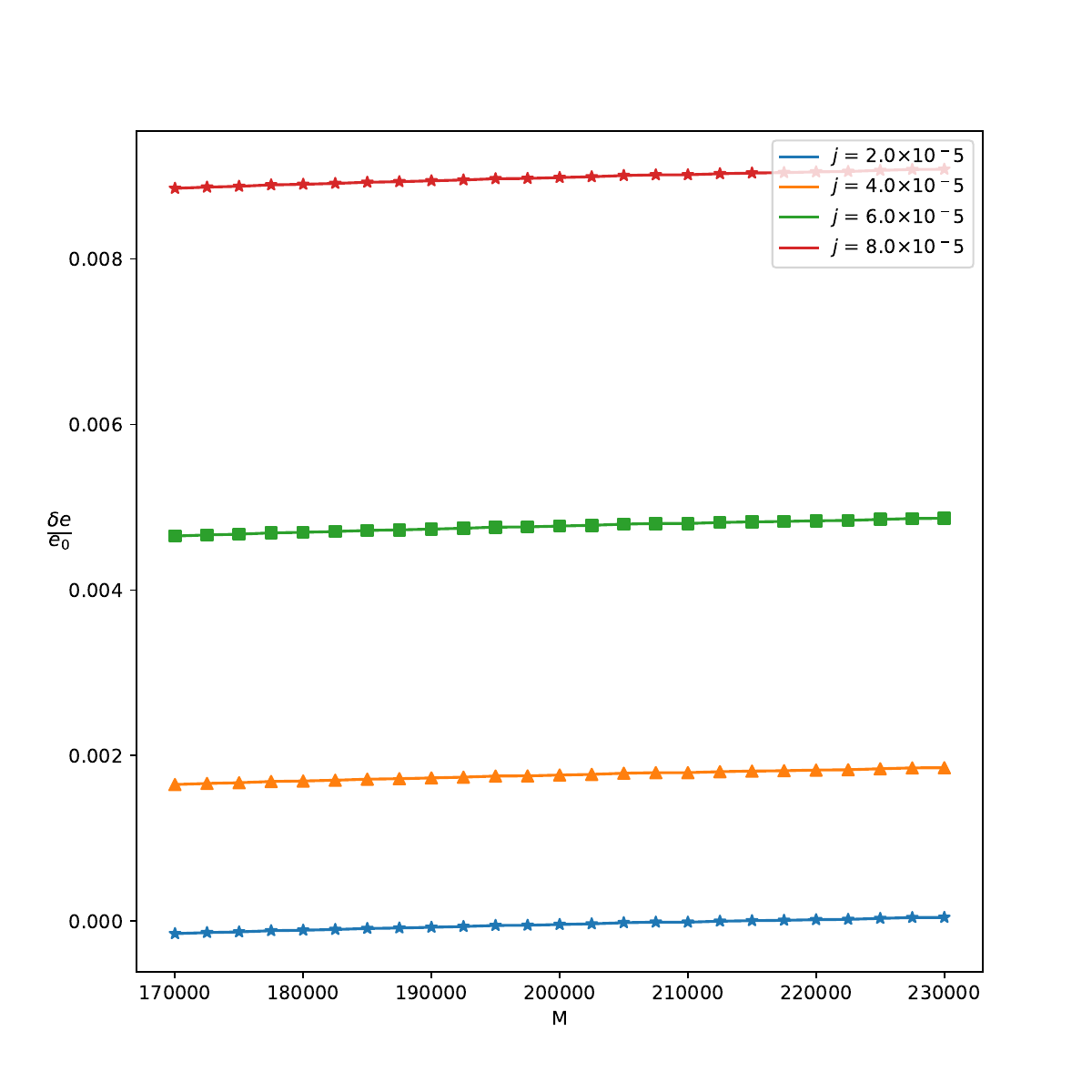}\\
\includegraphics[width=0.33\linewidth,height=0.3\linewidth]{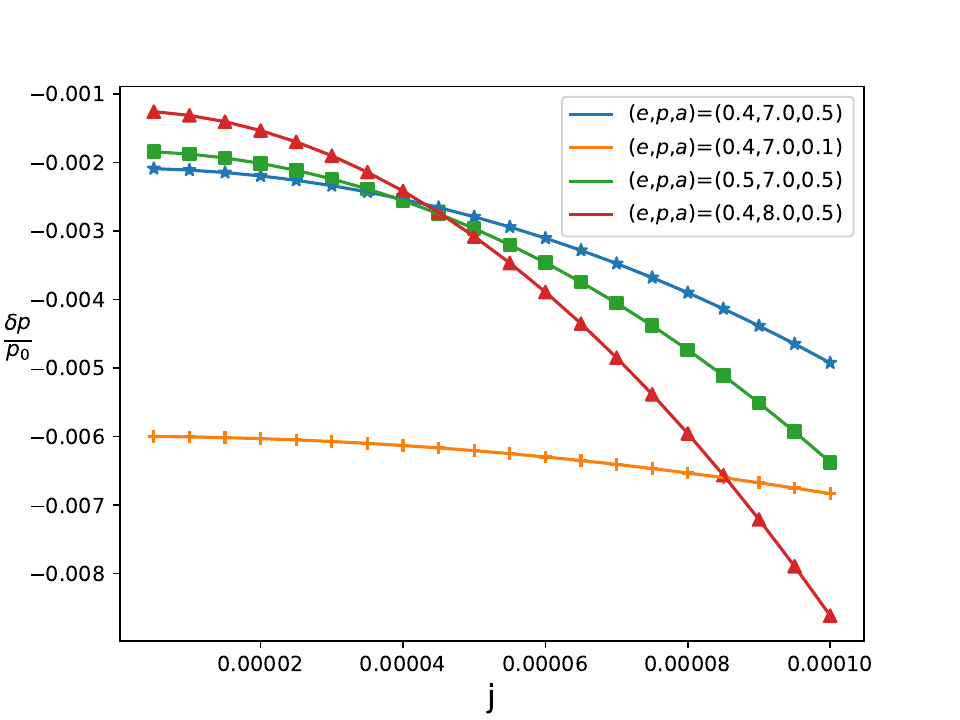}\includegraphics[width=0.332\linewidth,height=0.3\linewidth]{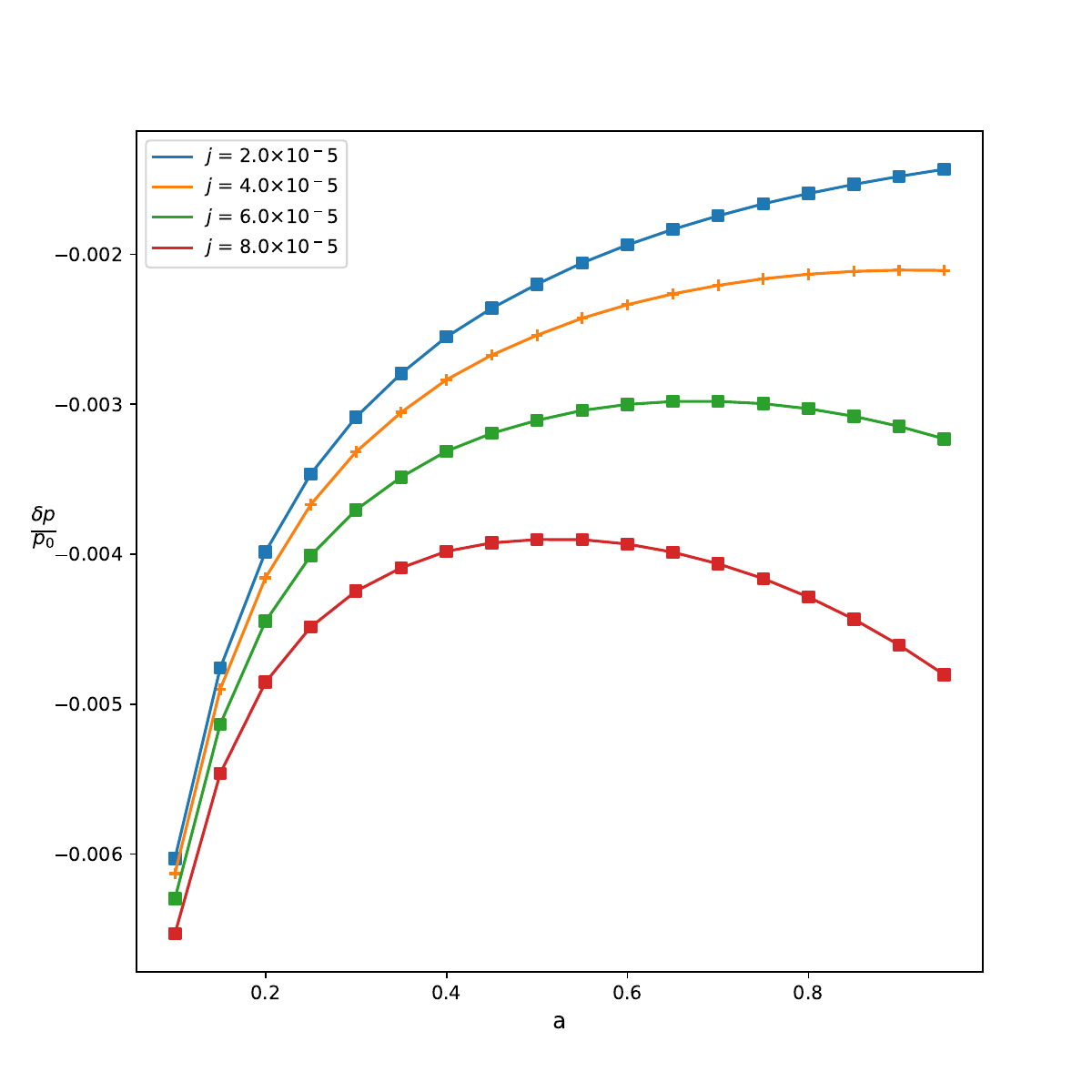} \includegraphics[width=0.33\linewidth,height=0.3\linewidth]{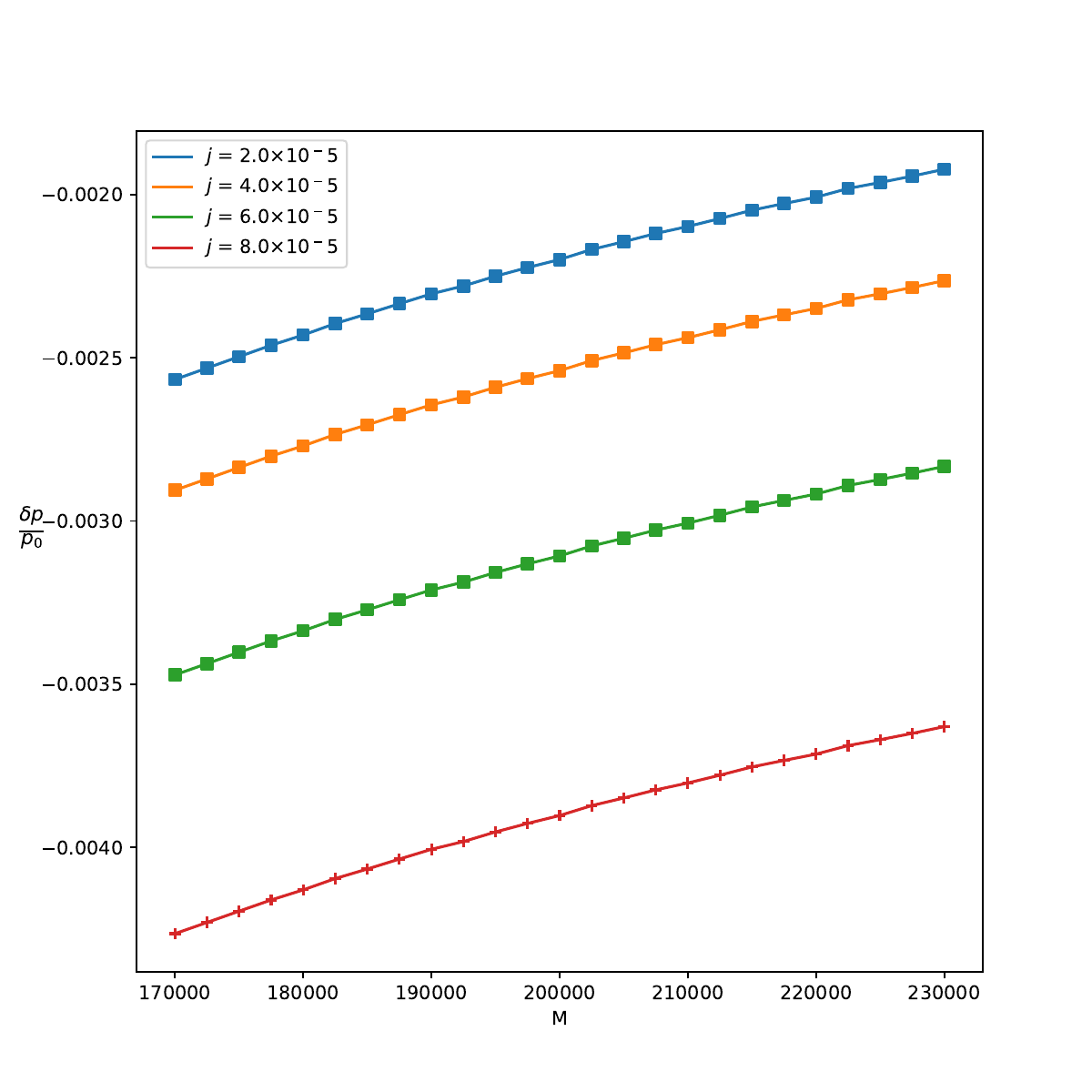}
	\end{minipage}
            \caption{Relation between relative varied orbit parameters for different black hole parameters when equating orbital frequencies, i.e., the
orbits of $(0, a, M, e_0, p_0)$ and $(j, a, M, e_0+ \delta e, p_0+ \delta p) $ have the same orbital frequencies. The black hole spin parameter is set to $a=0.5$ in the left and right panels, and the black hole mass parameter is set to $M=2\times10^{5}M_{\odot}$ in the left and middle panels. The orbit parameters are set to $e_0=0.4$ and $p_0=7.0$ in each panel.}
        \label{fig8}
\end{figure}

Let us to discuss the possible confusion of gravitational waves induced by the orbital parameters $e$ and $p$. As in ref.\cite{ref24}, one can calibrate the orbital frequency to look for combinations of parameters that might cause waveform confusion
\begin{equation}
    \begin{aligned}
        \Omega_{ki}(j=0,a_0,M_0, e_0,p_0)=\Omega_{si}(j,a_0,M_0, e_0+\delta e,p_0+\delta p),\quad\quad\quad i=r,\;\phi.
    \end{aligned}
\end{equation}
Here  $\delta p$ and $\delta e$  are two small deviations of the orbital parameters.
Solving above equations, we can  obtain deviations $\delta p$ and $\delta e$ and further discuss confusion problem arising from orbital parameters $e$ and $p$ for fixed $M$, $a$ and $j$. In Fig. \ref{fig8}, we present the relationship between the relative variations of orbital parameters and the black hole parameters $j$, $a$ and $M$ through matching the orbital frequency of Kerr and swirling-Kerr black holes. The results reveal that $\delta e/e_0$ increases rapidly and $\delta p/p_0$ decreases rapidly as $j$ increases, which means that changes of the swirling parameter $j$ must adjust the orbital parameters $e$ and $p$ to preserve consistent orbital frequencies. As the black hole spin parameter $a$ increases, $\delta e/e_0$ first decreases and then increases, whereas the change in $\delta p/p_0$ is exactly the opposite of that. This means that effects of the background rotation parameter $j$ on the relative variations $\delta e/e_0$ and $\delta p/p_0$ are distinctly different from those of the black hole spin parameter $a$.
With the increase in the black hole mass $M$, both $\delta e/e_0$ and $\delta p/p_0$ increase, but $\delta p/p_0$ has a higher rate of change.
Furthermore, we find that the change tendencies of $\delta e/e_0$ and $\delta p/p_0$ with the parameter $j$  are similar to those observed with the MOG parameter $\alpha$ in the STVG\cite{bd27a}, but the rates are different.
Through above analysis, we find that the confusion problem also exists in the swirling-Kerr case.

\section{summary}

It is important to study gravitational waves in the rotating black hole spacetime deviated from the usual Kerr one. In this paper, we have studied the EMRI gravitational waves induced by eccentric orbits in the equatorial plane within the framework of the swirling-Kerr black hole spacetimes. The swirling-Kerr black hole is a novel solution of vacuum general relativity and has an extra swirling parameter
characterizing the rotation of spacetime background. We find that the swirling parameter $j$ enhances the orbital differences between the swirling-Kerr black hole spacetime and the corresponding Kerr black hole spacetime, with effects opposite to those of the black hole spin parameter $a$. This distinction could serve as a potential probe for identifying the background's swirling.
As the parameter $p$ increases, the quantities $|\Delta\phi_{sK}-\Delta\phi_{K}|$ and $|T_{r(sK)}-T_{r(K)}|$ initially decrease and then increase, while the number of cycles $\mathcal{N}$ first increases and then decreases. On the other hand, as the orbital eccentricity  $e$ increases, the absolute values of $|\Delta\phi_{sK}-\Delta\phi_{K}|$ and $|T_{r(sK)}-T_{r(K)}|$ increase, while  the number of cycles $\mathcal{N}$ decreases.

For EMRI systems, the presence of the swirling parameter $j$ results in a phase delay in the gravitational waveforms, while the black hole spin parameter
$a$ suppresses the impact of the swirling parameter $j$ on the EMRI gravitational waves. As a result, extracting information about the swirling parameter
$j$  from the EMRI gravitational waves is much easier in the case of a slowly rotating black hole spacetime compared to a rapidly rotating black hole.
Finally, we also analyze the confusion problem of gravitational waves in the swirling-Kerr black hole spacetime. It is evident  that a high black hole spin results in a significant overlap between waveforms for different values of the swirling parameter $j$. For instance, in the case $a=0.5$, the overlap value for different $j$ remains high even at a retarded time $t=2 \times 10^6s$.  This significantly hinders our ability to detect the swirling parameter in the rapidly rotating black hole case by analyzing EMRI gravitational waveforms. Additionally, we explore the possible confusion of gravitational waves induced by the orbital parameters $e$ and $p$. For EMRI gravitational waves with the same orbital frequencies, the allowable relative variations of orbital parameters $\delta e/e_0$ increase rapidly, while $\delta p/p_0$  rapidly as $j$ increases. The change trends of $\delta e/e_0$ and $\delta p/p_0$ with the parameter $j$  are similar to those observed with the MOG parameter $\alpha$ in the STVG theory \cite{bd27a}, but the rates are different. As the black hole spin parameter $a$ increases, $\delta e/e_0$ first decreases and then increases, whereas the change in $\delta p/p_0$ is exactly the opposite of that. This means that effects of the background rotation parameter $j$ on the relative variations $\delta e/e_0$ and $\delta p/p_0$ are distinctly different from those of the black hole spin parameter $a$. With the increase in the black hole mass $M$, both $\delta e/e_0$ and $\delta p/p_0$ increase, but $\delta p/p_0$ has a higher rate of change.
These results provide further insight into the properties of EMRI gravitational waves and the background's swirling.

\section{\bf Acknowledgments}
This work was  supported by the National Natural Science
Foundation of China under Grant No.12275078, 11875026, 12035005, 2020YFC2201400 and the innovative research group of Hunan Province under Grant No. 2024JJ1006.

\end{document}